\documentclass[aps,prapplied,reprint,superscriptaddress]{revtex4-2}

\usepackage{graphicx}
\usepackage{physics}
\usepackage{hyperref}

\usepackage{amsmath}
\usepackage{amssymb} 
\usepackage{bm}
\usepackage{algorithm}
\usepackage{algpseudocode}

\algrenewcommand\algorithmicrequire{\textbf{Require:}}
\algrenewcommand\algorithmicensure{\textbf{Ensure:}}
\algrenewcommand\algorithmiccomment[1]{\hfill\(\triangleright\)~#1}

\begin{document}

\title{Divide-and-Conquer Neural Network Surrogates for Quantum Sampling: Accelerating Markov Chain Monte Carlo in Large-Scale Constrained Optimization Problems}

\author{Yuya Kawamata}
\email[Email: ]{u597035b@ecs.osaka-u.ac.jp}
\affiliation{Graduate School of Engineering Science, The University of Osaka,
1-3 Machikaneyama, Toyonaka, Osaka 560-8531, Japan}
\author{Yuichiro Nakano}
\affiliation{Graduate School of Engineering Science, The University of Osaka,
1-3 Machikaneyama, Toyonaka, Osaka 560-8531, Japan}
\author{Keisuke Fujii}
\affiliation{Graduate School of Engineering Science, The University of Osaka,
1-3 Machikaneyama, Toyonaka, Osaka 560-8531, Japan}
\affiliation{Center for Quantum Information and Quantum Biology, The University of Osaka, 1-2 Machikaneyama, Toyonaka, 560-0043, Japan.}
\affiliation{RIKEN Center for Quantum Computing (RQC), Hirosawa 2-1, Wako, Saitama 351-0198, Japan}
\affiliation{Graduate School of Informatics, Kyoto University, Sakyo-ku, Kyoto, 606-8501, Japan}

\date{\today}

\begin{abstract}
Quantum-enhanced Markov chain Monte Carlo (MCMC) [D.~Layden \mbox{\textit{et al.}},\href{https://www.nature.com/articles/s41586-023-06095-4}
{Nature 619, 282--287 (2023)}] is a promising way to use quantum-circuit samples as MCMC proposals to improve sampling efficiency.
On the other hand, many practical problems are large-scale and constrained, making it difficult to construct efficient proposal distributions in classical methods and slowing down MCMC mixing.
In this work, we propose a divide-and-conquer neural network surrogate framework for quantum sampling to accelerate MCMC under fixed Hamming weight constraints.
Our method divides the interaction graph for an Ising problem into subgraphs, generates samples using QAOA for those subproblems with an XY mixer, and trains neural network surrogates conditioned on the Hamming weight to provide proposal distributions for each subset while preserving the constraint.
In numerical experiments of Boltzmann sampling on 3-regular graphs, our method consistently accelerated mixing as the system size $N$ increased, with average improvements in the autocorrelation decay rate constant by speedup factors of about \mbox{$14.0$} and \mbox{$7.7$} over local and global Kawasaki dynamics, respectively.
We also applied the method to an MNIST feature mask optimization problem with $N=784$, obtaining faster energy convergence and a classification accuracy 2.03 percentage points higher than that of global Kawasaki dynamics after 50 MCMC steps.
These numerical results support the usefulness of surrogate proposals with a divide-and-conquer strategy for large-scale constrained sampling and finite-step optimization.
\end{abstract}

\maketitle

\section{INTRODUCTION}

Quantum computers are expected to become the next generation of computing devices, as they can solve certain problems faster than classical algorithms~\cite{shor1994algorithms,grover1997quantum, abrams1999quantum,harrow2009quantum}.
However, current Noisy Intermediate-Scale Quantum (NISQ) devices~\cite{preskill2018quantum} lack full quantum error correction, making it difficult to implement algorithms with theoretically guaranteed quantum speedup.
Nevertheless, as demonstrated by ``quantum computational supremacy"~\cite{arute2019quantum, zhong2020quantum, madsen2022quantum}, NISQ devices possess the potential to outperform classical computation, and significant efforts are underway to exploit this capability for practical applications. 
Variational Quantum Algorithms (VQAs) \cite{cerezo2021variational} are a representative approach, combining parameterized quantum circuits with classical optimization to search for solutions.
For instance, the Quantum Approximate Optimization Algorithm (QAOA)~\cite{farhi2014quantum} is a notable method for solving combinatorial optimization problems.
However, although VQAs have shown some success~\cite{mitarai2018quantum,farhi2018classification,peruzzo2014variational}, accurately evaluating Hamiltonian expectation values as objective functions remains challenging.
This is mainly due to the large number of measurements required as the circuit size increases and noise accumulates~\cite{gonthier2022measurements}, as well as the presence of barren plateaus in the optimization landscape \cite{mcclean2018barren}.

To explore useful applications of quantum-circuit samples, a key point to focus on is the ability to exploit an exponentially large Hilbert space for preparing quantum states and obtaining bit strings that are difficult for classical methods to produce through measurement.
In fact, random circuit sampling, used in demonstrations of quantum computational supremacy, is a direct application of this property~\cite{hangleiter2023computational}.
Based on this idea, a hybrid quantum–classical algorithm known as Quantum-Selected Configuration Interaction (QSCI)~\cite{kanno2023quantum} has been proposed.
QSCI uses samples from quantum circuits to identify important configurations, and then classically constructs and diagonalizes the Hamiltonian in the corresponding subspace, thereby enabling efficient approximation of the ground state.
Large-scale demonstrations on real quantum hardware have already been reported~\cite{robledo2025chemistry}, yielding accurate ground-state energies and chemically meaningful results from samples.

Another approach is to use the samples as a proposal distribution in Markov chain Monte Carlo (MCMC) \cite{layden2023quantum, nakano2025neural, nakano2024markov, lotshaw2023approximate, diez2023quantum}.
In the original approach of Layden et al.~\mbox{\cite{layden2023quantum}},
Trotterized quantum evolution from the current configuration generates symmetric proposals favoring small energy changes, without variational optimization of the circuit angles or direct approximation of a low-temperature Boltzmann distribution.
By contrast, QAOA-based approaches optimize circuits from a fixed initial state to generate low-energy samples for Metropolis--Hastings (MH) proposals, which are generally not symmetric~\mbox{\cite{nakano2025neural}}.
The underlying idea is that the distribution of samples from a circuit optimized by QAOA can approximate a low-temperature Boltzmann distribution, which is difficult to generate using classical methods \cite{diez2023quantum, lotshaw2023approximate}.
When these samples are used as proposals in the MH algorithm, they are accepted with high probability, enabling non-local transitions in the configuration space and consequently faster MCMC mixing.
This framework is applicable not only to low-temperature Boltzmann sampling but also to optimization problems that employ MCMC as a subroutine, such as simulated annealing, where it may lead to faster convergence to low-energy states.
On the other hand, running QAOA every time a proposal sample is needed in MCMC is computationally expensive and can limit practical use.
To overcome this issue, a method has been proposed to train a neural network known as Masked Autoencoder for Distribution Estimation (MADE) \cite{germain2015made, uria2016neural, larochelle2011neural} to approximate the sample distribution produced by quantum circuits \cite{nakano2025neural}.
This surrogate distribution can then be used as the proposal distribution, allowing efficient sampling without repeatedly executing quantum circuits and making the approach more practical.

However, previous studies have focused on small unconstrained problems, whereas practical problems are often large-scale and subject to constraints \cite{neuhaus20032d, nussbaumer2010free, nau2025quantum, mucke2023feature}.
In constrained sampling, the Markov-chain states must remain in the feasible set, so infeasible proposals must be excluded or rejected.
Even worse, as the problem size increases, each update typically changes only a small part of the configuration relative to the full state space.
As a result, constructing a proposal distribution that enables efficient exploration becomes more challenging as the system size increases.
An example of such a constraint is a fixed Hamming weight constraint.
This setting appears, for instance, in condensed matter physics problems such as droplet formation in lattice gases \cite{neuhaus20032d, nussbaumer2010free}, and in machine learning tasks such as constructing masks used for feature selection in images \cite{nau2025quantum, mucke2023feature}.
Under a fixed Hamming weight constraint, Kawasaki dynamics, also known as pair-flip updates, provide a proposal distribution that strictly satisfies the constraint by randomly selecting a pair of 0 and 1 and swapping them~\cite{hohenberg1977theory}.
However, it has been pointed out that updates based on such pair-flip dynamics may suffer from exponentially slow mixing with respect to the problem size, depending on the graph structure and the parameter regime \cite{kuchukova2025fast}.
To address this issue, we aim to construct a proposal distribution using samples generated from a quantum circuit, which can update more spins than pair-flip updates, while satisfying the constraint.

In this paper, we propose a divide-and-conquer neural network surrogate framework for quantum sampling to accelerate MCMC in large-scale constrained optimization problems.
More specifically, our method focuses on large-scale Ising problems under a fixed Hamming weight constraint.
The interaction graph for an Ising problem is divided into subgraphs, referred to as blocks.
For each block, samples are generated using a quantum circuit optimized by QAOA associated with 
smaller Ising problems defined on the subgraphs.
Specifically, we perform QAOA with an XY mixer~\cite{hadfield2019quantum, wang2020xy}, which preserves the Hamming weight, and the sample distribution is expected to approximate a constrained low-temperature Boltzmann distribution.
We then train a conditional MADE \cite{papamakarios2017masked} for each block, using the Hamming weight of the samples as a conditioning variable.
Each MADE conditioned on a Hamming weight can generate a surrogate distribution for the low-temperature Boltzmann distribution at that Hamming weight.
During MCMC, a block is randomly selected, and a proposal is drawn from the corresponding MADE while preserving the Hamming weight, which ultimately serves as a proposal distribution for simultaneously updating multiple spins within the block.

We perform numerical experiments to show that our method is effective.
First, we evaluate whether the proposal distribution in our method provides faster mixing than Kawasaki dynamics by studying constrained Boltzmann sampling on typical 3-regular graphs.
We fix the QAOA size to 16 qubits and increase the system size from $N=16$ to $N=512$.
We evaluate the mixing speed using the decay rate $\tau$ of the autocorrelation based on the Hamming distance.
Our method consistently achieves a speedup factor of about \mbox{$14.0$} over local Kawasaki dynamics, and a speedup factor of about \mbox{$7.7$} even over global Kawasaki dynamics.
These results indicate an approximately constant-factor improvement in the overlap-autocorrelation decay rate over the tested system-size range.
Then, we fix the system size to $N=256$ and increase the block size from 2 to 16 spins.
By comparing the mixing speed, we show that our method outperforms both variants of Kawasaki dynamics once the block size is at least 6, and that the mixing performance continues to improve as the block size increases.
Uniform block resampling using the same partitions instead exhibits slower decorrelation as the block size increases, supporting the benefit of energy-informed proposals beyond the larger update neighborhood alone.
Next, as a more practical application, we consider feature selection on MNIST under a fixed Hamming weight constraint~\cite{nau2025quantum, mucke2023feature}.
In this task, we optimize a QUBO with $N=784$ under the constraint to select pixels corresponding to important features.
Our method achieves faster energy convergence than Kawasaki dynamics. Furthermore, when MCMC is terminated after 50 steps, our method shows an improvement of about 2.03 percentage points in classification accuracy.

The remainder of this paper is organized as follows.
In Sec. \ref{sec:preliminaries}, we first review the preliminaries.
In Sec. \ref{sec:method}, we present our proposed method.
In Sec. \ref{sec:numerical}, we conduct numerical experiments to evaluate the effectiveness of the proposed method.
Finally, our conclusions are presented in Sec. \ref{sec:conclusion}.

\section{Preliminaries}
\label{sec:preliminaries}

\subsection{QAOA with XY mixer}\label{subsec:qaoa}

The Quantum Approximate Optimization Algorithm (QAOA) is a variational quantum algorithm that aims to produce bit strings $\bm{x}\in\{0,1\}^N$ with low values of the cost function $C(\bm x)$.
The cost function is assumed to be a quadratic binary objective of the form
\begin{align}
C(\bm{x})=\sum_{i<j} Q_{ij}x_i x_j+\sum_i q_i x_i+c,
\end{align}
where $Q_{ij}$, $q_i$, and $c$ are real coefficients.
Using the eigenvalue relation of the Pauli-$Z$ operator $Z_i\ket{\bm{x}}=(1-2x_i)\ket{\bm{x}}$, the cost Hamiltonian $H_{\rm C}$ can be explicitly defined as
\begin{align}
H_{\rm C}=\sum_{i<j}{J}_{ij} Z_i Z_j+\sum_i h_i Z_i+c',
\end{align}
with appropriately transformed coefficients $J_{ij}$, $h_i$, and $c'$.
The cost Hamiltonian is in Ising form and satisfies $H_{\rm C}\ket{\bm{x}}=C(\bm{x})\ket{\bm{x}}$.
QAOA uses a parameterized quantum circuit that alternates between a cost Hamiltonian $H_{\rm C}$ and a mixer Hamiltonian $H_{\rm M}$.
For depth $p$, the QAOA state is
\begin{align}
\ket{\psi(\bm{\gamma},\bm{\beta})}
=\Bigl(\prod_{\ell=1}^{p} e^{-i \beta_\ell H_{\rm M}}\, e^{-i \gamma_\ell H_{\rm C}}\Bigr)\ket{\psi_0}.
\end{align}
By measuring $\ket{\psi(\bm{\gamma},\bm{\beta})}$ in the computational basis, we obtain samples $\bm{x}$.
We optimize the parameters $(\bm{\gamma},\bm{\beta})$ on a classical computer so as to minimize the expected energy
\begin{align}
\mathcal{L}(\bm{\gamma},\bm{\beta})
=\langle \psi(\bm{\gamma},\bm{\beta})|H_{\rm C}|\psi(\bm{\gamma},\bm{\beta})\rangle,
\end{align}
which can be estimated from measured samples by evaluating $C(\bm{x})$.

To solve constrained problems, it is important that the state stays inside the feasible set.
We consider the fixed Hamming weight constraint in the optimization problem
\begin{align}
\sum_{i=1}^{N} x_i = K,
\end{align}
which means that exactly $K$ variables are equal to $1$.
To enforce this constraint, we use an XY mixer \cite{hadfield2019quantum, wang2020xy},
\begin{align}
H_{\rm M}
=\frac{1}{2}\sum_{(i,j) \in \mathcal{E}}\left(X_iX_j+Y_iY_j\right),
\end{align}
where $(i,j)\in \mathcal{E}$ are edges of the mixer graph.
The XY mixer swaps $\ket{01}\leftrightarrow\ket{10}$ along an edge, so it preserves the Hamming weight.
Therefore, if the initial state $\ket{\psi_0}$ is prepared in the Hamming weight $K$ subspace, then every QAOA layer preserves the constraint.

\subsection{Quantum-Enhanced MCMC via Neural Network Surrogate}\label{subsec:qaoa_mcmc}
We consider Boltzmann sampling for an energy function $E(\bm{x})$ over binary variables $\bm{x}\in\{0,1\}^N$ using MCMC.
The target distribution is the Boltzmann distribution
\begin{align}
\pi(\bm{x})
&=
\frac{1}{Z}e^{-\beta_{\pi} E(\bm{x})}, \\
Z
&=
\sum_{\bm x\in\{0,1\}^N} e^{-\beta_{\pi} E(\bm{x})} .
\label{eq:boltzmann}
\end{align}
In the Metropolis--Hastings (MH) algorithm \cite{metropolis1953equation, hastings1970monte}, given the current state $\bm{x}$, we draw a candidate $\bm{y}$ from a proposal distribution $q(\bm{y}\mid \bm{x})$ and accept it with probability
\begin{align}
\alpha(\bm{x}\to \bm{y})
&=
\min\!\left(1,\;
\frac{\pi(\bm{y})\,q(\bm{x}\mid \bm{y})}{\pi(\bm{x})\,q(\bm{y}\mid \bm{x})}
\right) \\
&=
\min\!\left(1,\;
e^{-\beta_{\pi}\{E(\bm{y})-E(\bm{x})\}}\frac{q(\bm{x}\mid \bm{y})}{q(\bm{y}\mid \bm{x})}
\right).
\label{eq:mh_accept}
\end{align}
If the move is accepted, we set the next state to $\bm{y}$; otherwise, we keep $\bm{x}$.
This accept-reject step ensures that $\pi$ is the stationary distribution of the Markov chain.

In the QAOA-based approach considered here, proposal construction is motivated by the observation that optimized QAOA samples can approximately follow a low-temperature Boltzmann distribution \mbox{\cite{nakano2025neural, lotshaw2023approximate, diez2023quantum}}.
Symmetric quantum proposal kernels avoid explicit evaluation of the proposal-probability ratio in the MH acceptance rule.
By contrast, the QAOA-generated proposals considered here are generally asymmetric, so their probabilities must be evaluated for the MH correction.
Also, to obtain samples from the exact distribution output by the quantum circuit, it was necessary to run the quantum circuit during the MCMC process.
To avoid these restrictions, the study in Ref.~\cite{nakano2025neural} replaces the quantum circuit by a Masked Autoencoder for Distribution Estimation (MADE) \cite{germain2015made, uria2016neural, larochelle2011neural, papamakarios2017masked} surrogate model trained on samples from the circuit.
MADE generates samples according to the autoregressive factorization
\begin{align}
\hat q(\bm x)
=
\prod_{i=1}^{N}\hat q\!\left(x_i \mid x_{<i}\right).
\label{eq:made_factorization}
\end{align}
This enables us to draw samples and compute the likelihood of any given sample, and we propose using a sample from MADE in the independent MH algorithm.
The acceptance probability becomes
\begin{align}
\alpha(\bm x \to \bm y)
=
\min\!\left(1,\;
e^{-\beta_{\pi}\{E(\bm y)-E(\bm x)\}}
\frac{\hat q(\bm x)}{\hat q(\bm y)}
\right).
\label{eq:mh_independent}
\end{align}
If the move is accepted, we set the next state to $\bm{y}$; otherwise, we keep $\bm{x}$.

\subsection{Kawasaki Dynamics}\label{subsec:constrained_mcmc}
We consider Boltzmann sampling under a fixed Hamming weight constraint.
We define the feasible set by
\begin{align}
\Omega_K
&=
\left\{
\bm x\in\{0,1\}^{N}
\ \middle|\ 
\sum_{i=1}^{N} x_i = K
\right\}.
\label{eq:Omega_k_method_sec2}
\end{align}
The target distribution is the Boltzmann distribution restricted to $\Omega_K$.
For $\bm{x}\in\Omega_K$, we define
\begin{align}
\pi_K(\bm{x})&=\frac{1}{Z_K} e^{-\beta_{\pi} E(\bm{x})},
\label{eq:target_distribution}
\\
Z_K
&=
\sum_{\bm{x}\in\Omega_K}e^{-\beta_{\pi} E(\bm{x})}.
\end{align}
When sampling under this constraint, a common approach is to use proposals that always satisfy the constraint, such as Kawasaki dynamics, also known as pair-flip moves or swap updates \cite{hohenberg1977theory,kawasaki1966diffusion}.
Kawasaki dynamics chooses a pair $(i,j)$ with $x_i=1$ and $x_j=0$, and proposes $\bm y$ obtained by swapping these two bits.
There are two standard variants of Kawasaki updates, which differ in how the $(1,0)$ pair is chosen.
In both cases, the move swaps a bit with value $1$ and a bit with value $0$, so the Hamming weight is preserved.
Global Kawasaki selects a pair of sites with opposite bits from the entire system:
\begin{align}
i &\sim \mathrm{Unif}\{m\in[N]: x_m=1\}, \\
j &\sim \mathrm{Unif}\{m\in[N]: x_m=0\},
\label{eq:kawasaki_global_choice}
\end{align}
and proposes $\bm y$ obtained by swapping $x_i$ and $x_j$.
Local Kawasaki selects only an adjacent pair with opposite bits.
We consider a graph $G=(V,\mathcal{E})$, where $V$ denotes the set of vertices and $\mathcal{E}$ denotes the set of edges of the graph $G$.
We choose an edge $(i,j)\in \mathcal{E}$ uniformly at random from $\mathcal{E}$.
If $x_i\neq x_j$, we propose $\bm y$ obtained by swapping $x_i$ and $x_j$; otherwise, we set $\bm y=\bm x$, resulting in a null move.
In both variants, the proposal kernel is symmetric.
Hence the proposal distribution cancels in the Metropolis--Hastings acceptance probability, and the acceptance rule reduces to the Metropolis form
\begin{align}
\alpha(\bm x\to \bm y)=\min\!\left(1,\;e^{-\beta_{\pi}\{E(\bm y)-E(\bm x)\}}\right).
\end{align}
If the move is accepted, we set the next state to $\bm{y}$; otherwise, we keep $\bm{x}$.

\section{Divide-and-Conquer Neural Network Surrogate Framework}\label{sec:method}

\begin{figure*}[t]
  \centering
  \includegraphics[width=\textwidth]{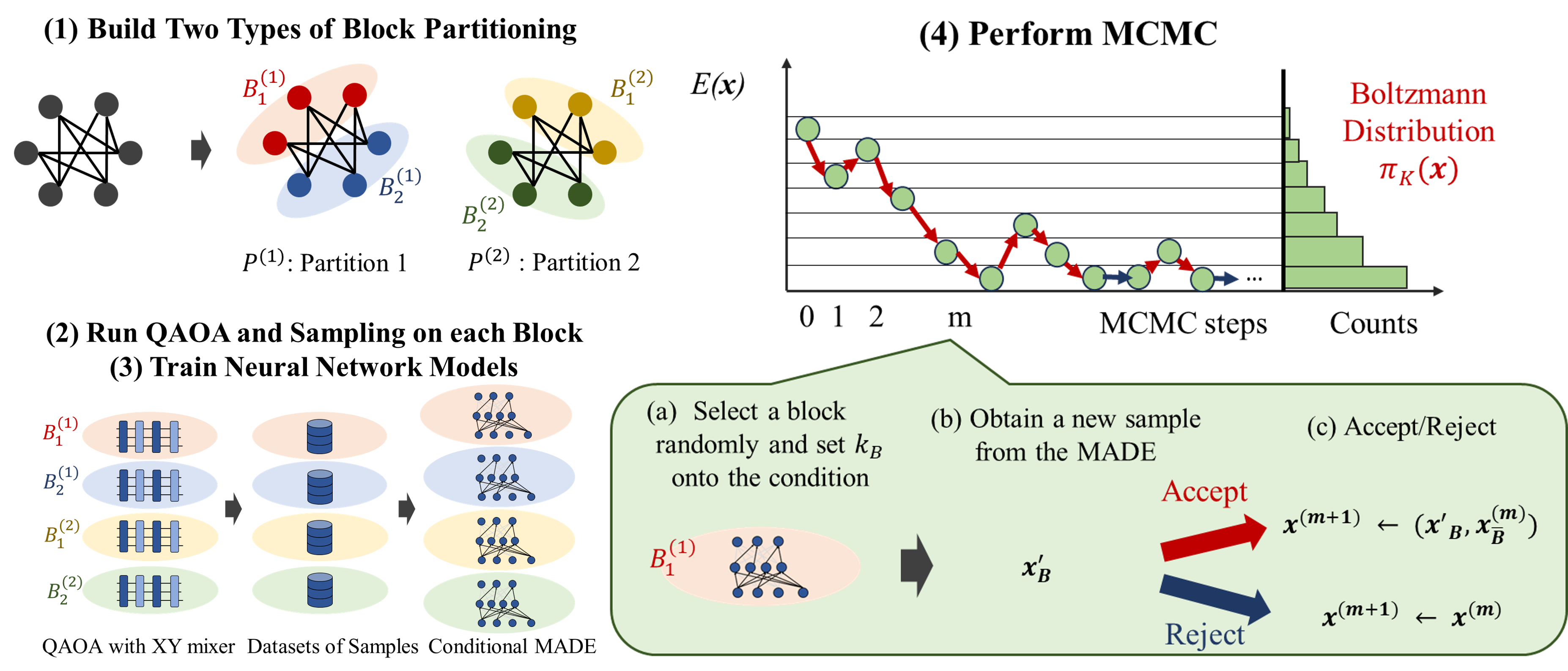}
  \caption{A schematic diagram of the proposed method in this study. The method comprises four main components and proceeds as follows: (1) Build two types of block partitioning. (2) Run QAOA and sampling on each block. (3) Train neural network models. (4) Perform MCMC.}
  \label{fig:dc_qaoa_mcmc}
\end{figure*}

We propose a divide-and-conquer neural network surrogate framework for quantum sampling to accelerate MCMC.
Here, we aim to sample from Eq.~(\ref{eq:target_distribution}).
Our method consists of four main steps, which are summarized in Fig.~\ref{fig:dc_qaoa_mcmc} and in Algorithm~\ref{alg:dc_qaoa_mcmc}. 
Here, ``divide-and-conquer'' refers to decomposing proposal construction into blockwise QAOA optimization and surrogate training, with the resulting models used in a global MCMC process rather than directly combining independently obtained block solutions.

\noindent\textbf{ (1) Build two types of block partitioning.}
Let $G$ be a graph with $N$ vertices.
For each $s\in\{1,2\}$, we construct sets of vertices:
\begin{align}
\mathcal{P}^{(s)}=\{B_1^{(s)},\dots,B_{M^{(s)}}^{(s)}\},
\end{align}
where each $B_m^{(s)}$ is a nonempty subset of vertices of $G$, indexed by $m$.
$B_m^{(s)}$ and $\mathcal{P}^{(s)}$ are referred to as a block and a partition, respectively.
These blocks form a partition of the vertex set of $G$, meaning that each vertex is contained in exactly one block in $\mathcal{P}^{(s)}$.

Each partition is constructed by a greedy procedure.
More specifically, we first choose a seed vertex and then repeatedly
add an unassigned neighboring vertex \mbox{$j$} that maximizes
\mbox{$\sum_{i\in B}|Q_{ij}|$} for the current block \mbox{$B$}, thereby favoring
strong interactions of either sign.
This is because we use a Hamiltonian that ignores connections to nodes outside the block in the subsequent QAOA step.
The larger the edge weights to nodes outside the block, the larger the deviation from the Hamiltonian corresponding to the target Boltzmann distribution.
The second partition $\mathcal{P}^{(2)}$ is generated in the same manner but with a different random initialization.
An important requirement is that the blocks in $\mathcal{P}^{(2)}$ should be sufficiently different to cross the boundaries of $\mathcal{P}^{(1)}$.
In particular, each block $B_m^{(2)}$ should intersect multiple blocks in $\mathcal{P}^{(1)}$.
This enables the Markov chain to make transitions that cross block boundaries.
A more detailed description of the partitioning algorithm is provided in Appendix~\ref{ap:partition}.

\noindent\textbf{ (2) Run QAOA and sampling on each block.}
We perform QAOA for each block using a Hamiltonian that includes only interactions within the block.
The target distribution in Eq.~\eqref{eq:target_distribution} is based on the following energy function:
\begin{align}
E(\bm{x})
&= \sum_{(i,j)\in\mathcal{E}} Q_{ij}\, x_i x_j \;+\; \sum_{i\in V} q_i\, x_i,
\label{eq:qubo_energy}
\end{align}
where $\bm x=(x_1,\ldots,x_N)\in\{0,1\}^N$, $V=\{1,\ldots,N\}$ is the set of variables, and $\mathcal{E}$ is the set of interacting pairs $(i,j)$ such that $Q_{ij}\neq 0$.
For each block $B_m^{(s)}$, we define the internal energy function of the block as:
\begin{align}
E_{B_m^{(s)}}\left(\bm {x}_{B_m^{(s)}}\right)
&=\sum_{(i,j)\in\mathcal{E}(B_m^{(s)})} Q_{ij} x_i x_j+\sum_{i\in B_m^{(s)}} q_i x_i,
\label{eq:block_cost_hamiltonian}
\end{align}
where $\bm x_{B_m^{(s)}}:= (x_i)_{i\in B_m^{(s)}} \in \{0,1\}^{|B_m^{(s)}|}$ denotes the restriction of $\bm x$ to the block $B_m^{(s)}$ and $\mathcal{E}(B_m^{(s)}):=\{(i,j)\in\mathcal{E}\mid i\in B_m^{(s)},\ j\in B_m^{(s)}\}$.
\begin{align}
H_{\rm C}^{\left(B_m^{(s)}\right)}\ket{\bm x_{B_m^{(s)}}}=E_{B_m^{(s)}}\left(\bm x_{B_m^{(s)}}\right)\ket{\bm x_{B_m^{(s)}}}.    
\end{align}
Importantly, this Hamiltonian does not include interactions between vertices in $B_m^{(s)}$ and those outside.
As a mixer, we use an XY mixer that conserves the Hamming weight within the block:
\begin{align}
H^{\left(B_m^{(s)}\right)}_{\mathrm{M}}
&=
\frac{1}{2}
\sum_{(i,j)\in \mathcal{E}_M(B_m^{(s)})}
\left(X_iX_j+Y_iY_j\right),
\label{eq:block_xy_mixer_internal}
\end{align}
where $\mathcal{E}_M(B_m^{(s)})$ is a chosen set of mixer edges inside $B_m^{(s)}$, and $X_i,Y_i$ are Pauli operators acting on qubit $i$.
The depth-$p$ block QAOA state is
\begin{align}
\ket{\psi_{B_m^{(s)}}(\boldsymbol{\gamma},\boldsymbol{\beta})}
&=
\Bigl(\prod_{\ell=1}^{p}
e^{-i\beta_\ell H^{\left(B_m^{(s)}\right)}_{\mathrm{M}}}
e^{-i\gamma_\ell H^{\left(B_m^{(s)}\right)}_{\rm C}}
\Bigr)\ket{\psi_{0,B_m^{(s)}}}.
\label{eq:block_qaoa_state_internal}
\end{align}
We optimize $(\boldsymbol{\gamma},\boldsymbol{\beta})$ by minimizing
\begin{align}
\mathcal{L}(\bm{\gamma},\bm{\beta})
=\langle \psi_{B_m^{(s)}}(\bm{\gamma},\bm{\beta})|H_{\rm C}^{\left(B_m^{(s)}\right)}|\psi_{B_m^{(s)}}(\bm{\gamma},\bm{\beta})\rangle.
\end{align}
As a simple choice, we use the uniform initial state $\ket{\psi_{0,B_m^{(s)}}}=\ket{+}^{\otimes |B_m^{(s)}|}$ and perform the optimization once per block.
The initial state has a binomial distribution of the block Hamming
weight \mbox{$k_{B_m^{(s)}}$}, concentrated near \mbox{$|B_m^{(s)}|/2$}.
This distribution over Hamming-weight sectors is preserved by
the block QAOA circuit.
Within these subspaces, the state can approach a low-temperature Boltzmann distribution.
In contrast, subspaces with other Hamming weights typically have smaller initial amplitudes and are therefore less effectively optimized.
When $K$ is small and the Hamming weight within each block tends to be small, one can instead prepare an initial state whose expected Hamming weight in each block is smaller by applying a small-angle rotation gate to each qubit once.
We also note that prior work suggests that even fixed angles can still accelerate sampling, and therefore useful performance can be expected for other Hamming weights as well \cite{nakano2025neural}.
When generating training data, we run the circuit from multiple initial states to obtain diverse samples.

\noindent\textbf{ (3) Train neural network models.}

\begin{figure}[t]
  \centering
  \includegraphics[width=0.8\linewidth]{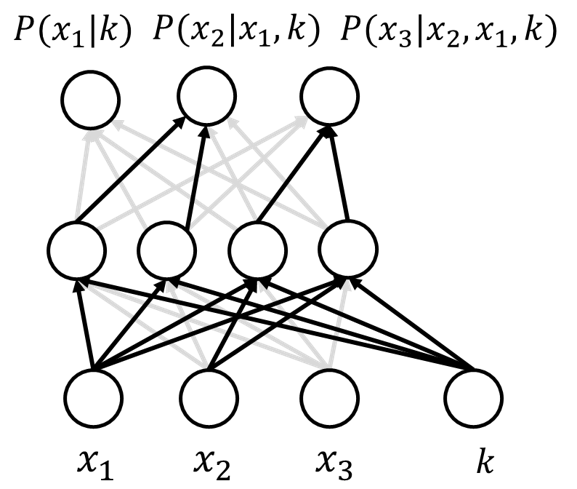}
  \caption{A conceptual diagram of the conditional MADE. Thick lines indicate active connections, while thin lines indicate absent connections. The context \mbox{$k$} denotes the block Hamming weight \mbox{$k_B$}, and the outputs represent \mbox{$P(x_1\mid k)$}, \mbox{$P(x_2\mid x_1,k)$}, and \mbox{$P(x_3\mid x_1,x_2,k)$}.}
  \label{fig:cmade}
\end{figure}

We train a conditional MADE \cite{papamakarios2017masked} for each block, using samples generated by a circuit optimized by QAOA with an XY mixer.
Figure~\ref{fig:cmade} illustrates the conditional MADE architecture.
The masks ensure that each output factor \mbox{$P(x_i\mid x_{<i},k)$} depends only on the preceding variables \mbox{$x_{<i}$} and the conditioning variable \mbox{$k$}, rather than on \mbox{$x_i$} or subsequent variables.
The key difference from the standard MADE is that the context $k$ is provided to all hidden layers.
In our implementation, \mbox{$k_B$} is one-hot encoded and concatenated
with the block configuration at the input.
The context units are assigned autoregressive degree zero, allowing the conditioning information to propagate through the masked network.
During training, we feed the block Hamming weight of each QAOA sample together with the sample itself.
At sampling time, by specifying a desired block Hamming weight, the model is expected to produce samples that mimic QAOA outputs conditioned on that Hamming weight.

Given a conditioning value $k_{B_m^{(s)}}\in\{0,1,\ldots,|B_m^{(s)}|\}$, the conditional MADE defines an autoregressive
proposal distribution over $\bm x_{B_m^{(s)}}\in\{0,1\}^{|B_m^{(s)}|}$:
\begin{align}
\hat q_{\theta}^{(B_m^{(s)})}&\left(\bm x_{B_m^{(s)}} \mid k_{B_m^{(s)}}\right) \notag \\
&=
\prod_{t=1}^{|B_m^{(s)}|}
\hat q_{\theta}^{(B_m^{(s)})}\!\left(\left(x_{B_m^{(s)}}\right)_{i_t}\,\middle|\,\left(x_{B_m^{(s)}}\right)_{i_{<t}},\,k_{B_m^{(s)}}\right),
\label{eq:cond_made_factor_kB}
\end{align}
where $o_{B_m^{(s)}}=(i_1,\ldots,i_{|B_m^{(s)}|})$ is a fixed ordering of the variables in block $B_m^{(s)}$, $i_{<t}:=(i_1,\ldots,i_{t-1})$, and $\theta$ denotes the parameters of the neural network.
We train the model by maximum likelihood using QAOA block samples
$\{\bm x_{B_m^{(s)}}^{(n)}\}_{n=1}^{N_{B_m^{(s)}}}$, where $N_{B_m^{(s)}}$ is the total number of samples on $B_m^{(s)}$, together with their block Hamming weights
\begin{align}
k_{B_m^{(s)}}^{(n)}:=\sum_{i\in B_m^{(s)}}\left(\bm x_{B_m^{(s)}}^{(n)}\right)_i,
\label{eq:block_weight_training}
\end{align}
namely,
\begin{align}
\max_{\theta}\;
\sum_{n=1}^{N_{B_m^{(s)}}}
\log \hat q_{\theta}^{\left(B_m^{(s)}\right)}\!\left(\bm x_{B_m^{(s)}}^{(n)}\mid k_{B_m^{(s)}}^{(n)}\right).
\label{eq:cond_made_mle_kB}
\end{align}

\noindent\textbf{ (4) Perform MCMC.}
At each MCMC step, we choose a partition type $s\in\{1,2\}$ uniformly at random, and then choose a block index $m$ uniformly at random from $\{1,\ldots,M^{(s)}\}$, defining the selected block as $B:=B_m^{(s)}$.
Because the global Hamming weight is fixed to $K$, the allowed block Hamming weight is uniquely
determined by the current complement $\bm x_{\bar B}$:
\begin{align}
k_B
&=
K-\sum_{i\in\bar B} x_i .
\label{eq:block_weight_method_sec2}
\end{align}
Using this $k_B$ as the context, we draw a block proposal $\bm x'_B$ from the conditional MADE and explicitly obtain the probability $\hat q^{(B)}_\theta(\bm x'_{B} \mid k_B)$.
If $\bm x'_B$ does not satisfy $\sum_{i\in B} (\bm x'_B)_i = k_B$, we immediately reject it.
Otherwise, we set $\bm y =(\bm x'_B, \bm x_{\bar B})$, and accept $\bm y$ with the MH acceptance probability
\begin{align}
\alpha(\bm x\to \bm y)
&=
\min\!\left(
1,\;
e^{-\beta_{\pi}\{E(\bm y)-E(\bm x)\}}
\frac{\hat q^{(B)}_\theta(\bm x_B \mid k_B)}{\hat q^{(B)}_\theta(\bm x'_B \mid k_B)}
\right).
\label{eq:block_mh_accept_internal}
\end{align}
If the move is accepted, we set the next state to $\bm{y}$; otherwise, we keep $\bm{x}$.
This transition kernel leaves the target distribution $\pi_K$ in Eq.~\eqref{eq:target_distribution} invariant.

\begin{algorithm}[H]
\caption{Divide-and-conquer neural network surrogate framework for quantum sampling}
\label{alg:dc_qaoa_mcmc}
\begin{algorithmic}
\Statex \textbf{Input:} Target energy $E(\bm x)$, inverse temperature $\beta_{\pi}$, total Hamming weight $K$, number of MCMC steps $T$.
\Statex \textbf{Output:} A Markov chain $\{\bm{x}^{(t)}\}_{t=0}^{T}$ with stationary distribution $\pi_K$.

\Statex
\Statex \textbf{(1) Build two types of block partitioning}
\State Construct two vertex partitions $\mathcal{P}^{(1)},\mathcal{P}^{(2)}$.

\Statex
\Statex \textbf{(2) Run QAOA and sampling on each block}
\For{$s=1,2$}
  \For{$m=1,\ldots,M^{(s)}$}
    \State Run QAOA on block $B_m^{(s)}$ with an XY mixer.
    \State Store the samples in $D_{s,m}$.
  \EndFor
\EndFor

\Statex
\Statex \textbf{(3) Train neural network models}
\For{$s=1,2$}
  \For{$m=1,\ldots,M^{(s)}$}
    \State Train a conditional MADE for block $B_m^{(s)}$ using $D_{s,m}$.
  \EndFor
\EndFor

\Statex
\Statex \textbf{(4) Perform MCMC}
\State Initialize $\bm x^{(0)}\in\Omega_K$.
\For{$t=0,1,\dots,T-1$}
  \State Select $s\in\{1,2\}$ uniformly at random.
  \State Select $m\in\{1,\ldots,M^{(s)}\}$ uniformly at random.
  \State Set $B\gets B_m^{(s)}$.
  \State Compute $k_B \gets K-\sum_{i\in \bar B} x_i^{(t)}$.
  \State Draw $\bm x'_B$ from MADE for $(s,m)$ conditioned on $k_B$.
  \If{\mbox{$\sum_{i\in B} (x'_B)_i \neq k_B$}}
    \State $\bm x^{(t+1)}\gets \bm x^{(t)}$.   (Reject)
  \Else
    \State Set $\bm y=(\bm x'_B,\bm x_{\bar B}^{(t)})$.
    \State $\alpha\gets \min\!\left(1,\;e^{-\beta_{\pi}\{E(\bm y)-E(\bm x^{(t)})\}}
    \dfrac{\hat q^{(B)}_\theta(\bm x_B^{(t)}\mid k_B)}{\hat q^{(B)}_\theta(\bm x'_B \mid k_B)}\right)$.
    \If{$\alpha \ge \mathrm{Uniform}(0,1)$}
      \State $\bm x^{(t+1)}\gets \bm y$.  (Accept)
    \Else
      \State $\bm x^{(t+1)}\gets \bm x^{(t)}$.  (Reject)
    \EndIf
  \EndIf
\EndFor
\end{algorithmic}
\end{algorithm}

\section{Numerical Experiments}\label{sec:numerical}

In this section, we present numerical experiments to show the effectiveness of our method.
Our method provides energy-informed proposal distributions in which multiple spins can be updated at once while preserving the constraint, with the aim of enabling faster mixing than pair-flip updates.
To assess this, we study constrained Boltzmann sampling on 3-regular graphs at the calibrated target inverse temperature \mbox{$\beta_{\pi}=0.5$}.
We measure the decay of the overlap autocorrelation in two settings: increasing the system size with a fixed block size, and increasing the block size with a fixed system size.
Next, as a concrete example, we consider an application to feature mask construction for MNIST.
This problem can be formulated as an Ising optimization problem with a fixed Hamming weight constraint.
We compare the performance of our method in terms of energy convergence and classification accuracy when the process is stopped after a finite number of steps.

\subsection{3-regular graph}\label{sec:3regular}

We consider a Boltzmann sampling problem on a 3-regular graph with $N$ vertices and random couplings.
For each edge $(i,j)$, the coupling is drawn independently from a Gaussian distribution $Q_{ij} \sim \mathcal{N}(0,1)$, and no linear terms are included.
We impose the fixed Hamming weight constraint $\sum_{i=1}^{N} x_i = N/2$.
In our method, the system is partitioned into blocks of nominal size \mbox{$|B|$}; if \mbox{$N$} is not divisible by this size, the remaining vertices form one smaller block.
As an additional baseline, we also consider uniform block resampling using exactly the same block partitions and block-selection procedure as in the proposed method.
For a selected block \mbox{$B$} with current block Hamming weight \mbox{$k_B$}, the candidate configuration is drawn uniformly from the \mbox{$\binom{|B|}{k_B}$} configurations having exactly the same block Hamming weight.
In our implementation, these configurations are explicitly enumerated.
The proposal is symmetric and is accepted using the Metropolis rule based on the full-system energy difference.
Thus, this baseline has the same constraint-preserving block-update neighborhood as the proposed method, but its proposal distribution does not incorporate energy information.
The comparison therefore tests the benefit of the energy-informed QAOA--MADE proposal beyond enlarging the update neighborhood alone.
These simulations do not take into account noise in actual quantum hardware.

We use the decay rate $\tau$ to quantify the convergence rate, with larger $\tau$ indicating faster MCMC convergence.
First, we introduce an overlap for autocorrelation, corresponding to the Hamming distance \cite{hukushima1996exchange, katzgraber2006universality}.
Specifically, we run two independent MCMC chains, $\bm x(t)$ and $\bm x'(t)$, and define the overlap at time $t$ in the representation
$x_i(t)\in\{0,1\}$ as
\begin{align}
q(t)
&=
\frac{1}{N}\sum_{i=1}^{N} x_i(t)\,x'_i(t).
\label{eq:overlap_q}
\end{align}
We compute the autocorrelation function of $q(t)$ as
\begin{align}
\rho(l)
&:=
\frac{\mathbb{E}\!\left[(q(t)-\bar q)(q(t+l)-\bar q)\right]}
{\mathbb{E}\!\left[(q(t)-\bar q)^2\right]},
&
\bar q
&=
\mathbb{E}[q(t)],
\label{eq:autocorr_q}
\end{align}
where $l$ denotes the lag.
A fast decay of the overlap autocorrelation indicates active exploration of the configuration space and is therefore widely used as a proxy for fast mixing of other observables, including the energy.
In particular, since our focus is on improving the efficiency of exploration in the configuration space, this metric is appropriate for our study.
We set the target inverse temperature to \mbox{$\beta_{\pi}=0.5$}.
This target inverse temperature was chosen to approximately match the representative effective inverse temperature of the QAOA sample distributions.
The scaling results reported below therefore demonstrate faster decorrelation over the tested system-size range in this calibrated temperature regime; the same scaling behavior or speedup need not hold at arbitrary target inverse temperatures.
The calibration is described in Appendix~\mbox{\ref{ap:3regular}}, and the QAOA and MADE settings are summarized in Appendix~\mbox{\ref{ap:qaoa_made_settings}}.

\begin{figure}[t]
  \centering
  \includegraphics[width=1.0\linewidth]{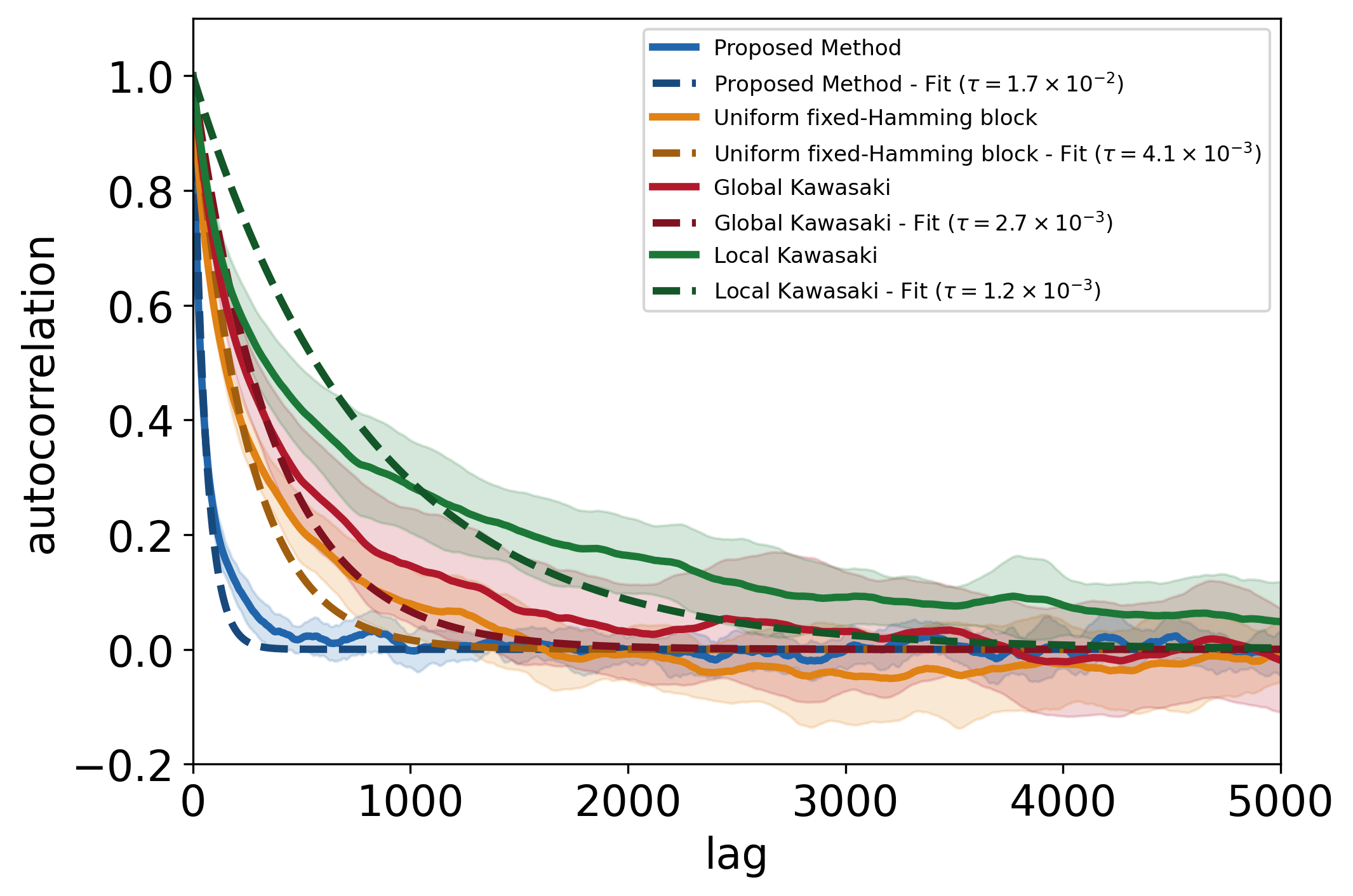}
  \caption{Autocorrelation for \mbox{$N=256$} with block size $|B|=16$. The yellow curve denotes uniform block resampling using the same partitions as the proposed method. The solid curves show the mean over 12 independent replica-pair autocorrelation functions, and the shaded regions show one standard deviation across the replica pairs at each lag. The dashed curves show fits of the form \mbox{$\exp(-\tau l)$}.}
  \label{fig:3regular_example}
\end{figure}

Figure~\mbox{\ref{fig:3regular_example}} shows an example of the MCMC results for
\mbox{$N=256$} and block size \mbox{$|B|=16$}.
The horizontal axis represents the lag $l$, and the vertical axis shows the autocorrelation $\rho(l)$.
The blue solid line corresponds to our proposed method, while the red and green solid lines represent global and local Kawasaki dynamics, respectively.
The yellow solid line shows the uniform block-resampling baseline described above.
The autocorrelation of our method decays faster, indicating faster MCMC mixing.
The dashed lines show fits of the form \mbox{$\exp(-\tau l)$}, and we use $\tau$ as a measure of the decay rate.
The comparison with uniform block resampling directly illustrates the benefit of the energy-informed QAOA--MADE proposal beyond the use of a larger block-update neighborhood alone.
Details of the fitting, uncertainty estimation, and acceptance-rate analysis are provided in Appendix~\mbox{\ref{ap:tau_fitting}} and Appendix~\mbox{\ref{ap:3regular}}.

\begin{figure}[t]
  \centering
  \includegraphics[width=1.0\linewidth]{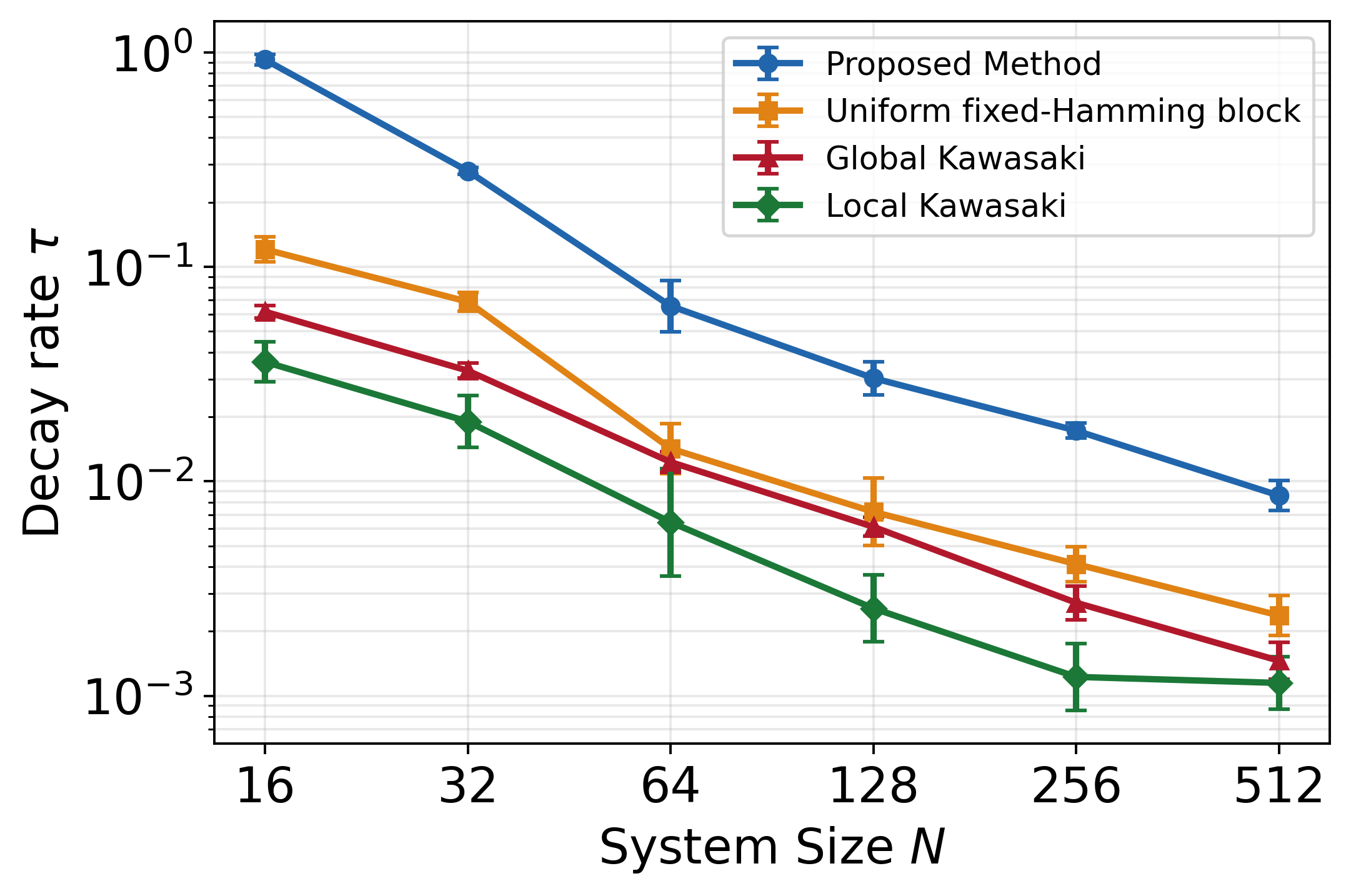}
  \caption{The dependence on the system size $N$ for 3-regular graphs. The block size is fixed as \mbox{$|B|=16$}. The yellow curve denotes uniform block resampling using the same partitions as the proposed method. The vertical axis shows the decay rate $\tau$, with larger values corresponding to more rapid convergence. The error bars show the 95\% confidence intervals obtained using the delete-one-replica-pair jackknife described in Appendix~\mbox{\ref{ap:tau_fitting}}.}
  \label{fig:3regular_system}
\end{figure}

First, we examine the dependence on the system size while keeping the block size fixed.
The block size is set to $|B|=16$, and the system size $N$ is increased.
The result is shown in Fig.~\ref{fig:3regular_system}.
Since a larger value of $\tau$ corresponds to faster convergence, our method consistently exhibits faster convergence than Kawasaki dynamics even as the system size increases.
Averaged over different system sizes, the ratios of $\tau$ show that our method achieves a speedup factor of \mbox{$7.7$} over global Kawasaki dynamics and \mbox{$14.0$} over local Kawasaki dynamics.
These results are consistent with an approximately constant-factor improvement in the decay rate over the tested system-size range; they do not establish more favorable asymptotic scaling with \mbox{$N$}.
The uniform block-resampling baseline, which uses the same partitions and block-update size but no energy-informed proposal distribution, is also included for comparison.
Its substantially slower decorrelation shows that the larger update neighborhood alone does not account for the performance of the proposed method.
The acceptance-rate analysis in Appendix~\mbox{\ref{ap:3regular}} further shows that the acceptance behavior does not systematically deteriorate as the total system size increases when the block size is fixed.

\begin{figure}[t]
  \centering
  \includegraphics[width=1.0\linewidth]{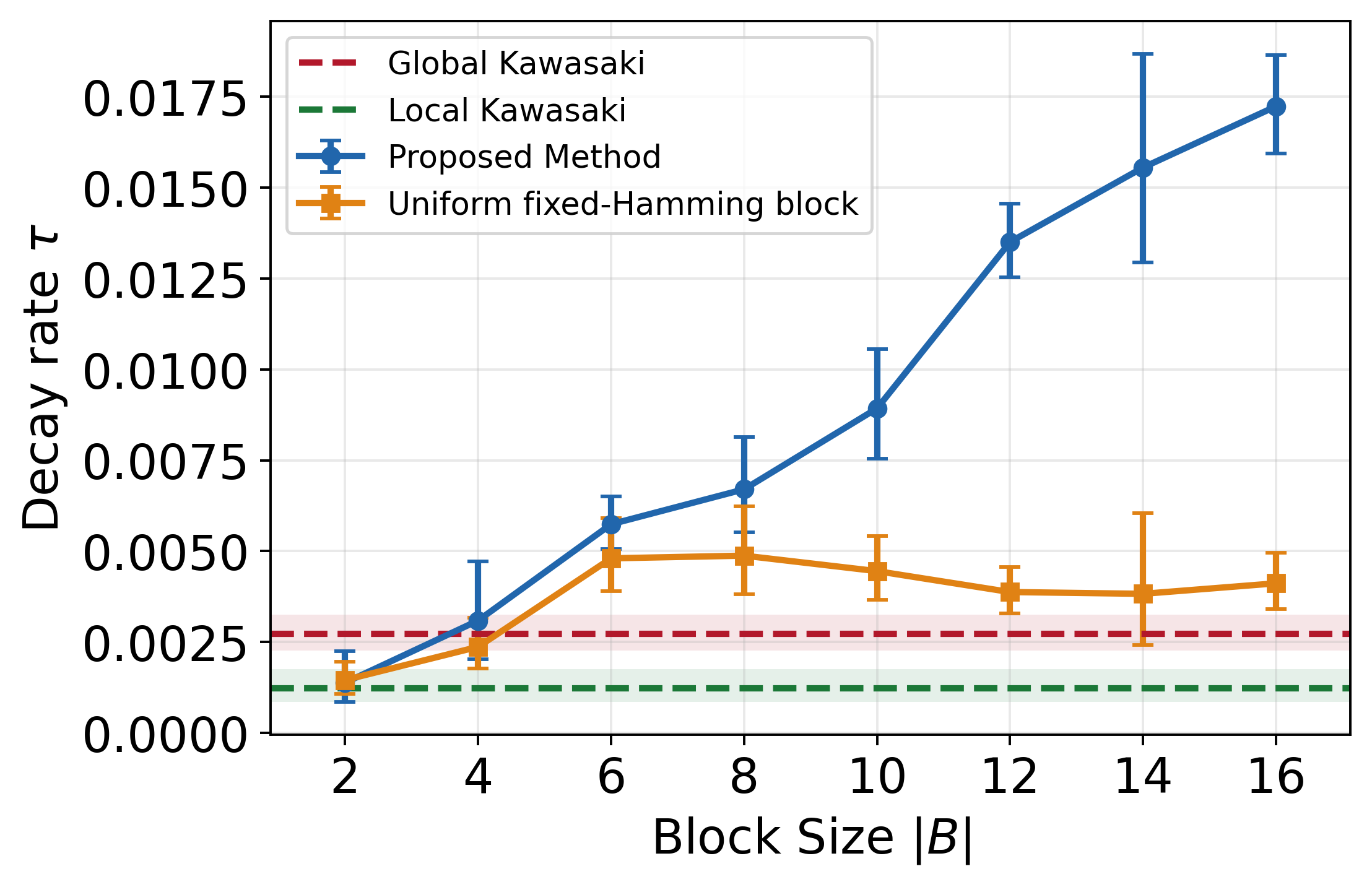}
  \caption{The dependence on the nominal block size $|B|$ for a 3-regular graph with \mbox{$N=256$}. The yellow curve denotes uniform block resampling using the same partitions as the proposed method. The vertical axis shows the decay rate $\tau$, with larger values corresponding to more rapid convergence. The error bars on the proposed method results show the 95\% jackknife confidence intervals. Global and local Kawasaki dynamics are independent of \mbox{$|B|$} and are shown as dashed horizontal lines.}
  \label{fig:3regular_block}
\end{figure}

Next, we investigate the dependence on the block size while keeping the system size fixed.
The system size is set to $N=256$, and the block size $|B|$ is increased.
The result is shown in Fig.~\ref{fig:3regular_block}.
Our method consistently outperforms both variants of Kawasaki dynamics when $|B| \geq 6$.
Importantly, the uniform block-resampling baseline shows the opposite trend:
its decay rate \mbox{$\tau$} decreases as the block size increases.
This demonstrates that simply enlarging the update neighborhood does not by itself
improve the mixing performance.
Rather, the comparison supports the benefit of combining larger block updates with
energy-informed QAOA--MADE proposals that capture useful internal-energy structure.
This is consistent with the Boltzmann-like character of the proposals and is further
supported by the acceptance-rate analysis in Appendix~\mbox{\ref{ap:3regular}},
which shows a weaker rejection penalty than for uniform resampling as the block size increases.

\subsection{Application: MNIST Feature Selection}\label{sec:mnist}

We study a concrete MCMC optimization application under a fixed Hamming weight constraint: feature selection from images~\cite{nau2025quantum, mucke2023feature, peng2005feature}.
The task is to construct a binary mask that selects only $K$ pixels from an image as features.
More specifically, we solve the QUBO constructed from the training data under a fixed Hamming weight constraint.
For the resulting variables $x_i \in \{0,1\}$, the corresponding pixel is selected as a feature if $x_i = 1$. In this formulation, the number of selected features, $K$, is directly represented by the fixed Hamming weight.
The motivation for selecting exactly $K$ pixels using a binary mask is to ensure explainability, unlike in methods such as principal component analysis.
For more details on the construction of the QUBO, please refer to Appendix~\ref{ap:mnist}.

\begin{figure}[t]
  \centering
  \includegraphics[width=1.0\linewidth]{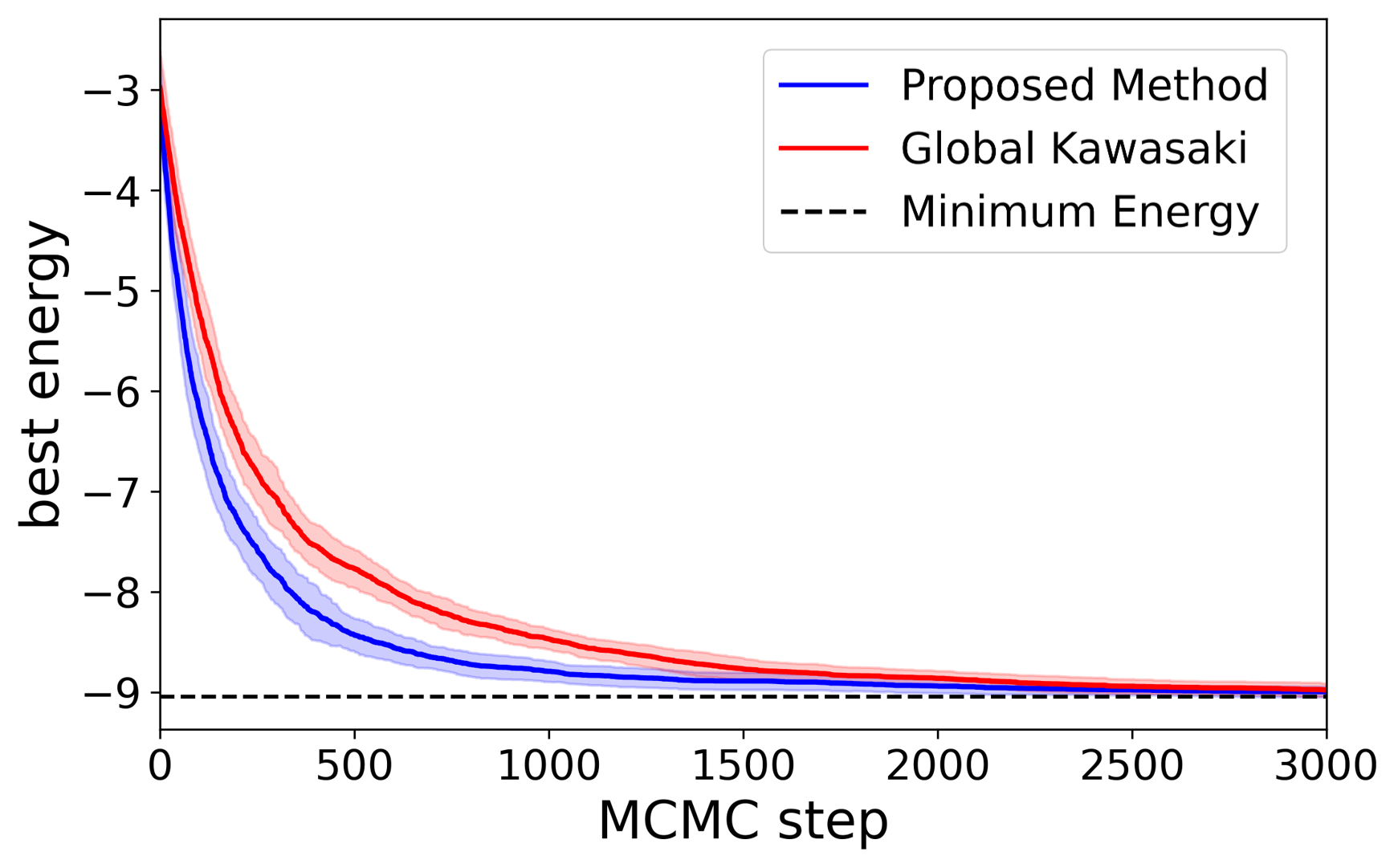}
    \caption{MCMC results for MNIST feature selection.
    The horizontal axis shows the number of MCMC steps, and the vertical
    axis shows the best-so-far energy.
    The blue and red curves show the mean best-so-far energies across
    replicas for the proposed method and global Kawasaki dynamics,
    respectively.
    The shaded regions indicate \mbox{$\pm1$} population standard deviation across 20 replicas
    at each step, with a divisor of 20 in the variance calculation.
    The black dashed line denotes the best-found energy obtained by
    global Kawasaki dynamics across the tested runs and inverse
    temperatures within the MCMC step budgets used in this experiment.
    This reference value is not a certified constrained ground-state
    energy.}
  \label{fig:mnist_energy}
\end{figure}

\begin{table*}[t]
  \centering
  \setlength{\tabcolsep}{14pt}
  \renewcommand{\arraystretch}{1.35}
  \begin{tabular}{|l|c|c|}
    \hline
      & \textbf{\shortstack{Accuracy\\After 50 steps}}
      & \textbf{\shortstack{Accuracy\\After 3000 steps}} \\
    \hline
    \textbf{Proposed Method}   & 79.51 \% & 80.51 \% \\
    \hline
    \textbf{Global Kawasaki} & 77.48 \% & 80.50 \% \\
    \hline
    \textbf{Linear Terms 50}   & \multicolumn{2}{c|}{76.03 \% } \\
    \hline
    \textbf{Random 50}         & \multicolumn{2}{c|}{71.00 \% } \\
    \hline
  \end{tabular}
  \caption{Classification accuracy on MNIST with the mask applied. 
  Linear Terms 50 denotes a mask constructed by selecting 50 pixels using only the linear part of the QUBO, whereas Random 50 denotes a mask constructed by randomly selecting 50 pixels. 
  For Proposed Method and Global Kawasaki, the reported accuracies are obtained by terminating MCMC after 50 and 3000 steps.}
  \label{tab:accuracy_rates}
\end{table*}

In our numerical experiments, we use MNIST images of size $28\times 28$ and perform feature selection from $N=784$ pixel features with $K=50$.
We extract a $50$-dimensional feature vector and train a logistic regression classifier; we compare methods by classification accuracy.
For the proposed method, we divide the variables into $50$ blocks, with the QAOA block sizes ranging from $12$ to $16$.
Unlike the uniform superposition obtained by applying Hadamard gates to all qubits, we prepare an initial state biased so that the Hamming weight is concentrated around $k_{B}=2$ with small-angle single-qubit rotation gates, and then start the optimization from that neighborhood.
As a baseline, we use simulated annealing with Kawasaki pair-flip updates, which preserves the total Hamming weight.
This problem is sparse and contains isolated vertices, so local Kawasaki is not an appropriate baseline.
While simulated annealing is often run with a cooling schedule from high temperature to low temperature to avoid local minima, in this experiment we prioritize fast convergence and start the search directly at a low temperature corresponding to $\beta_{\pi}=100$.
The inverse temperature is held fixed throughout each run, without a cooling schedule.
We also present the results for other values of $\beta_{\pi}$ in Appendix~\ref{ap:mnist} for reference.

Figure~\ref{fig:mnist_energy} shows the results.
The horizontal axis is the number of MCMC steps, and the vertical axis is the lowest energy found up to that step.
Compared with global Kawasaki, our proposed method reaches lower energy solutions in fewer steps.
In this application, MCMC is used as a heuristic optimizer, and our aim is to evaluate finite-step search performance rather than to validate a sampling distribution or certify global optimality.
The black dashed line is an empirical best-found reference obtained from global Kawasaki runs at several inverse temperatures within the MCMC step budgets used here.
Although the energy trajectories stabilize and the resulting masks improve classification performance, these observations do not establish that this reference is the true constrained ground-state energy.
Table~\ref{tab:accuracy_rates} reports the classification accuracy of logistic regression trained on the selected features.
Here, Random 50 denotes the accuracy obtained by selecting 50 pixels uniformly at random from the 784 pixels, averaged over 10 random masks.
Linear Terms 50 denotes the accuracy obtained by selecting 50 pixels using only the linear part of the QUBO.
These two baselines do not perform optimization.
Therefore, our proposed method, similarly to global Kawasaki, achieves higher accuracy than both Random 50 and Linear Terms 50, demonstrating practical performance for feature selection under the fixed Hamming weight constraint.
For the proposed method and global Kawasaki, we also evaluate the accuracy when the MCMC run is stopped at step $50$ and at step $3000$.
At step $3000$, the energies reached by the two methods are already similar, so the classification accuracies are also similar.
In contrast, when we stop at step $50$, there is still a clear energy gap, and the proposed method achieves better accuracy, where the accuracy gap is 2.03 percentage points.
Overall, these results demonstrate improved search performance under
a limited MCMC step budget.
The proposed method obtains useful low-energy feature masks in fewer steps, whereas the classification accuracies of the two methods are nearly identical after \mbox{$3000$} steps.
The observed advantage therefore concerns finite-step search efficiency, rather than evidence of a better final optimum.

\section{Conclusion}\label{sec:conclusion}

In this paper, we proposed a divide-and-conquer neural network surrogate framework for quantum sampling to accelerate MCMC in large-scale constrained optimization problems. 
Recent studies have shown that quantum-enhanced Markov chain Monte Carlo can use quantum samples as a proposal distribution to accelerate convergence to a target distribution. 
However, many practical problems are large-scale and constrained, and in such settings it remains difficult to construct efficient proposal distributions in classical methods, often resulting in slow MCMC mixing.
To address this challenge, we presented the method for MCMC under fixed Hamming weight constraints. 
Our method divides the interaction graph for an Ising problem into subgraphs (i.e., blocks), generates samples using QAOA with an XY mixer, and trains neural networks conditioned on the Hamming weight to provide proposal distributions for each block while preserving the constraint. 

In numerical experiments of Boltzmann sampling on 3-regular graphs, our method consistently accelerated mixing as the system size $N$ increased, with average improvements in the autocorrelation decay rate by speedup factors of about \mbox{$14.0$} and \mbox{$7.7$} over local and global Kawasaki dynamics, respectively.
We also applied the method to an MNIST feature mask optimization problem with $N=784$, obtaining faster energy convergence and a classification accuracy 2.03 percentage points higher than that of global Kawasaki dynamics after 50 MCMC steps.
Overall, the framework offers a useful way to exploit QAOA-generated
samples by training reusable classical proposal distributions
for large-scale constrained MCMC.
Its robustness to hardware noise and the scaling of the total
quantum and classical resource requirements, including QAOA
optimization, circuit sampling, and MADE training, remain to be
assessed.

\begin{acknowledgments}
This work is supported by MEXT Quantum Leap Flagship Program (MEXT Q-LEAP) Grant No. JPMXS0120319794, JST COI-NEXT Grant No. JPMJPF2014, and JST CREST JPMJCR24I3.
\end{acknowledgments}

\appendix

\section{Algorithm for Partitioning}\label{ap:partition}

We build two block partitions by a greedy procedure on the weighted graph, summarized in Algorithm~\ref{alg:block_partition}.
For each partition $\mathcal{P}^{(s)}$, we start from unassigned seed vertices and grow each block $B_m^{(s)}$ until it reaches the prescribed block size.
At each step, we add an unassigned neighboring vertex \mbox{$j$} maximizing
\mbox{$\sum_{i\in B_m^{(s)}}|Q_{ij}|$}.
If no unassigned neighbor exists, we add a random unassigned vertex instead.
The second partition is generated in the same manner with a different random initialization.
This procedure tends to place strongly coupled vertices in the same block, while producing two different decompositions of the graph.
Using both partitions can enable Hamming-weight transfer across
block boundaries, although the extent of this transfer depends on
their cross-block connectivity.
If the two partitions do not sufficiently cross each other’s boundaries, it is necessary to introduce an additional procedure, such as swapping vertices that remain in the same block, in order to obtain more distinct block decompositions.
For \mbox{$N=256$} and \mbox{$|B|=2$}, we explicitly constructed the second partition so that the two vertices in each of its blocks belong to different blocks of the first partition.

\begin{algorithm}[H]
\caption{Construction of two block partitions}
\label{alg:block_partition}
\begin{algorithmic}
\Statex \textbf{Input:}  Graph $G=(V,\mathcal{E})$ with couplings, block sizes $\mathrm{Blocksize}(s,m)$
\Statex \textbf{Output:}  $\mathcal{P}^{(1)}=\{B_{1}^{(1)},\dots,B_{M^{(1)}}^{(1)}\}$ and $\mathcal{P}^{(2)}=\{B_1^{(2)},\dots,B_{M^{(2)}}^{(2)}\}$

\For{$s=1,2$}
    \State Initialize the seed and mark vertices unassigned
    \For{$m=1,\dots,M^{(s)}$}
        \State Choose an unassigned seed vertex and add it to $B_m^{(s)}$
        \While{$|B_m^{(s)}|< \mathrm{Blocksize}(s,m)$}
            \State Let $\mathcal{C}$ be unassigned neighbors of $B_m^{(s)}$
            \If{$\mathcal{C}\neq\emptyset$}
                \State Add \mbox{$j^\star\in\arg\max_{j\in\mathcal{C}}
                \sum_{i\in B_m^{(s)}}|Q_{ij}|$} to $B_m^{(s)}$
            \Else
                \State Add a random unassigned vertex to $B_m^{(s)}$
            \EndIf
        \EndWhile
    \EndFor
\EndFor

\State \Return $\mathcal{P}^{(1)}$ and $\mathcal{P}^{(2)}$
\end{algorithmic}
\end{algorithm}

Because the same greedy objective is used to construct both partitions, different random initializations do not necessarily guarantee sufficiently distinct block boundaries.
It is therefore useful to monitor the complementarity of the resulting partitions and, when the overlap is too large, to modify the second partition, for example by swapping vertices between its blocks while preserving the prescribed block sizes.
In the 3-regular-graph experiments, we quantified this complementarity by defining
\begin{align}
D_{\mathrm{pair}}
=
1-
\frac{
\displaystyle\sum_{m,n}
\binom{\left|B_m^{(1)}\cap B_n^{(2)}\right|}{2}
}{
\displaystyle\sum_m
\binom{\left|B_m^{(1)}\right|}{2}
},
\label{eq:partition_pair_difference}
\end{align}
which is the fraction of unordered vertex pairs placed in the same block by
\mbox{$\mathcal{P}^{(1)}$} but in different blocks by \mbox{$\mathcal{P}^{(2)}$}.
Because the two partitions have the same block-size profile in our
experiments, the same value is obtained when their roles are reversed.
The measured values are summarized in
Table~\mbox{\ref{tab:partition_pair_difference}}.
For the representative setting \mbox{$N=256$} and \mbox{$|B|=16$} used in Fig.~\mbox{\ref{fig:3regular_energy_histogram}}(c), we obtained \mbox{$D_{\mathrm{pair}}=0.550$}.
Thus, \mbox{$55.0\%$} of the within-block vertex pairs in the first partition are separated by the second partition.
This value quantifies the difference between the two partitions at the vertex-pair level, but does not by itself establish that all additional block-weight invariants are removed.
The energy-distribution and autocorrelation checks in Appendix~\mbox{\ref{ap:3regular}} provide complementary empirical evidence for the proposed sampler evaluated there; they do not isolate the effect of partition complementarity alone.
Thus, \mbox{$D_{\mathrm{pair}}$} can serve as a practical diagnostic for constructing
the two partitions, although its appropriate threshold may depend on the
problem instance and insufficiently distinct partitions should be regenerated
or modified by vertex swaps.

\begin{table}[t]
  \centering
  \caption{
  Fraction of vertex pairs placed in the same block by
  \mbox{$\mathcal{P}^{(1)}$} but in different blocks by
  \mbox{$\mathcal{P}^{(2)}$} in the 3-regular-graph experiments.
  The setting \mbox{$(N,|B|)=(256,16)$} is shared by the system-size and
  block-size sweeps. For \mbox{$N=|B|=16$}, the entire system forms a single
  block, and hence \mbox{$D_{\mathrm{pair}}=0$} trivially.
  The dash for \mbox{$(N,|B|)=(256,2)$} marks the explicitly
  constructed cross-partition pairing described in the text.
  }
  \label{tab:partition_pair_difference}
  \begin{ruledtabular}
    \begin{tabular}{ccc}
      \mbox{$N$} & \mbox{$|B|$} & \mbox{$D_{\mathrm{pair}}$} (\%) \\
      \hline
       16 & 16 &  0.0 \\
       32 & 16 & 52.5 \\
       64 & 16 & 35.0 \\
      128 & 16 & 41.8 \\
      256 & 16 & 55.0 \\
      512 & 16 & 54.9 \\
      \hline
      256 &  2 & -- \\
      256 &  4 & 57.6 \\
      256 &  6 & 59.7 \\
      256 &  8 & 56.6 \\
      256 & 10 & 48.4 \\
      256 & 12 & 46.3 \\
      256 & 14 & 48.5 \\
    \end{tabular}
  \end{ruledtabular}
\end{table}

\section{Biased initial states and block-Hamming-weight-sector coverage}
\label{ap:biased_initial_states}

For a block \mbox{$B$} of size \mbox{$b=|B|$}, we prepare the product state
\begin{align}
\ket{\phi_B(r)}
&{}=
\bigotimes_{i\in B}R_y\!\left(\theta(r)\right)\ket{0}_i,
&
\theta(r)&{}=2\arcsin\sqrt{r},
\label{eq:biased_product_state}
\end{align}
where \mbox{$r$} is the probability of measuring a given qubit in state
\mbox{$\ket{1}$}.  The resulting block Hamming weight \mbox{$k_B$} follows
\begin{align}
P_b(k_B\mid r)
=\binom{b}{k_B}r^{k_B}(1-r)^{b-k_B},
\qquad
\mathbb{E}[k_B]=br.
\label{eq:biased_binomial_distribution}
\end{align}

The rotation parameter is chosen from the characteristic number of occupied
variables expected in a block.  For a global constraint of Hamming weight
\mbox{$K$} and \mbox{$M$} blocks of comparable size, this number is approximately
\mbox{$\bar{k}_B=K/M$}; more generally, for a block of size \mbox{$b$}, it is
\mbox{$\bar{k}_B\simeq bK/N$}.  After selecting a representative integer sector
\mbox{$k_B^*$} near this value, we choose
\begin{align}
r_B^*=\frac{k_B^*}{b},
\qquad
\theta_B^*=2\arcsin\sqrt{\frac{k_B^*}{b}}.
\label{eq:biased_rotation_choice}
\end{align}
Indeed, \mbox{$r_B^*=k_B^*/b$} maximizes \mbox{$P_b(k_B^*\mid r)$} with respect to
\mbox{$r$}, while also giving \mbox{$\mathbb{E}[k_B]=k_B^*$}.  We use the same
prescription for every block, with the numerical angle adjusted to its size.
In the MNIST experiment, we used \mbox{$k_B^*=2$} for \mbox{$b=12$}--\mbox{$16$}, corresponding
to \mbox{$\theta_B^*=0.8411$}--\mbox{$0.7227$} rad; although \mbox{$bK/N=0.77$}--\mbox{$1.02$},
this choice retains \mbox{$26.9$}--\mbox{$27.0\%$} of the probability at \mbox{$k_B=1$} while
placing \mbox{$75.2$}--\mbox{$76.3\%$} in the neighboring low-weight sectors
\mbox{$1\leq k_B\leq3$}.

For the generation of training samples, using only one value of \mbox{$r$} would
leave the training set strongly biased toward a narrow range of sectors.
We therefore keep the optimized QAOA angles fixed, repeat the circuit for a
set \mbox{$\mathcal{R}$} of rotation parameters, and pool the same number of
measurements from each value.  The resulting sector distribution is
\begin{align}
P_b^{\rm(train)}(k_B)
=\frac{1}{|\mathcal{R}|}
\sum_{r\in\mathcal{R}}
\binom{b}{k_B}r^{k_B}(1-r)^{b-k_B},
\label{eq:training_sector_mixture}
\end{align}
which smooths the sector imbalance of a single binomial distribution.  In
the 3-regular-graph experiments, we used
\mbox{$\mathcal{R}=\{0.1,0.3,0.5,0.7,0.9\}$} and collected \mbox{$50{,}000$}
measurements for each value of \mbox{$r$}.  In the MNIST experiment, we used
\mbox{$\mathcal{R}=\{0.1,0.2,\ldots,0.9\}$} and collected \mbox{$30{,}000$}
measurements for each value; for \mbox{$b=12$}--\mbox{$16$}, this gives approximately
\mbox{$6.5\times10^3$}--\mbox{$2.31\times10^4$} raw samples per Hamming-weight sector
and per block in expectation.

\section{Autocorrelation fitting and uncertainty estimation}
\label{ap:tau_fitting}

For each setting in the 3-regular-graph overlap-autocorrelation analysis, we generate \mbox{\(R=12\)} independent pairs
of MCMC trajectories and calculate the overlap autocorrelation
\mbox{\(\rho_r(l)\)} for each replica pair \mbox{\(r\)}.
The solid curves in the autocorrelation plots show the pair average
\begin{align}
\overline{\rho}(l)
=
\frac{1}{R}\sum_{r=1}^{R}\rho_r(l),
\end{align}
while the shaded regions show one standard deviation across the
replica pairs.
Thus, the shaded regions represent the pair-to-pair variation of the
autocorrelation curves and are not confidence intervals for the fitted
decay rate.

We fit the averaged autocorrelation using
\begin{align}
\rho_{\mathrm{fit}}(l)=\exp(-\tau l),
\end{align}
where the amplitude is fixed to unity because
\mbox{\(\rho(0)=1\)}.
The fit is performed over \mbox{\(1\leq l\leq5000\)}, excluding non-finite points and retaining only points satisfying \mbox{\(0<\overline{\rho}(l)\leq1\)}.
Weighted nonlinear least squares is used, with the standard deviation \mbox{\(\sigma_l\)} across replica pairs corresponding to a weight proportional to \mbox{\(1/\sigma_l^2\)}.
Here, \mbox{\(\tau\)} is the decay rate in inverse MCMC steps, so a larger
value of \mbox{\(\tau\)} indicates faster decorrelation.

The uncertainty of \mbox{\(\tau\)} is estimated using a delete-one-pair
jackknife.
Specifically, we omit replica pair \mbox{\(r\)}, recompute the averaged
autocorrelation and its standard deviation from the remaining
\mbox{\(R-1\)} pairs, and repeat the complete fit to obtain
\mbox{\(\widehat{\tau}_{(-r)}\)}.
Since \mbox{\(\tau\)} is positive, the jackknife calculation is performed for
\mbox{\(\theta_{(-r)}=\log\widehat{\tau}_{(-r)}\)}.
The corresponding standard error is
\begin{align}
\mathrm{SE}_{\mathrm{JK}}
=
\sqrt{
\frac{R-1}{R}
\sum_{r=1}^{R}
\left[
\theta_{(-r)}-\overline{\theta}_{(-)}
\right]^2
},
\end{align}
where
\mbox{\(\overline{\theta}_{(-)}
=R^{-1}\sum_r\theta_{(-r)}\)}.
The reported \mbox{\(95\%\)} confidence interval is
\begin{align}
\left[
\exp\!\left(
\log\widehat{\tau}
-
t_{R-1,0.975}\mathrm{SE}_{\mathrm{JK}}
\right),
\,
\exp\!\left(
\log\widehat{\tau}
+
t_{R-1,0.975}\mathrm{SE}_{\mathrm{JK}}
\right)
\right].
\end{align}
The error bars in Fig.~\mbox{\ref{fig:3regular_system}} and those on the proposed-method points in Fig.~\mbox{\ref{fig:3regular_block}} show these jackknife confidence intervals. For the block-size dependence, the Kawasaki results are independent of the block size and are therefore shown as dashed horizontal lines without shaded confidence bands.

\section{QAOA and MADE settings}
\label{ap:qaoa_made_settings}

Table~\mbox{\ref{tab:qaoa_made_settings}} summarizes the principal
hyperparameters used in the 3-regular-graph experiments.
All sample counts are given per block.

\begin{table}[t]
\caption{QAOA and MADE settings.}
\label{tab:qaoa_made_settings}
\centering
\footnotesize
\begin{ruledtabular}
\begin{tabular}{@{}p{0.34\columnwidth}p{0.58\columnwidth}@{}}

\multicolumn{2}{c}{\textit{QAOA}}\\
\hline
Depth
& \mbox{$p=5$} \\

Optimization shots
& \mbox{$0$} (exact state-vector expectation) \\

Mixer
& Periodic-ring XY mixer \\

Initial state
& \mbox{$\lvert+\rangle^{\otimes |B|}$} \\

Optimizer
& Two-point stochastic FD with Adam \\

Optimization schedule
& \mbox{$p=1\!\to\!5$}; 100 steps/depth; 3 restarts at \mbox{$p=1$};
100-step polish at \mbox{$p=5$} \\

Training measurements
& \mbox{$250{,}000$} \\

Measurement breakdown
& \mbox{$r=\{0.1,0.3,0.5,0.7,0.9\}$};
\mbox{$50{,}000$} samples per value of \mbox{$r$}; \mbox{$\sigma_{\rm jitter}=0.05$} \\

\hline
\multicolumn{2}{c}{\textit{Conditional MADE}}\\
\hline

QAOA data
& \mbox{$250{,}000$} \\

Train/validation
& \mbox{$225{,}000/25{,}000$} \\

Hidden layers
& \mbox{$2\times256$} (GELU) \\

Optimizer
& Adam \\

Learning rate
& \mbox{$2.0\times10^{-3}$} \\

Epochs
& 10 \\

Batch size
& 1024 \\

\end{tabular}
\end{ruledtabular}
\end{table}

For each sample \mbox{$\bm{x}_B$}, the block Hamming weight
\mbox{$k_B=\sum_{i\in B}x_i$} was encoded as a \mbox{$(|B|+1)$}-dimensional
one-hot vector and concatenated with \mbox{$\bm{x}_B$} at the MADE input.
The context units were assigned autoregressive degree zero, allowing
a single conditional MADE to represent all Hamming-weight sectors
of each block.

\section{3-regular graph}\label{ap:3regular}

\begin{figure}[t]
  \centering
  \includegraphics[width=1.0\linewidth]{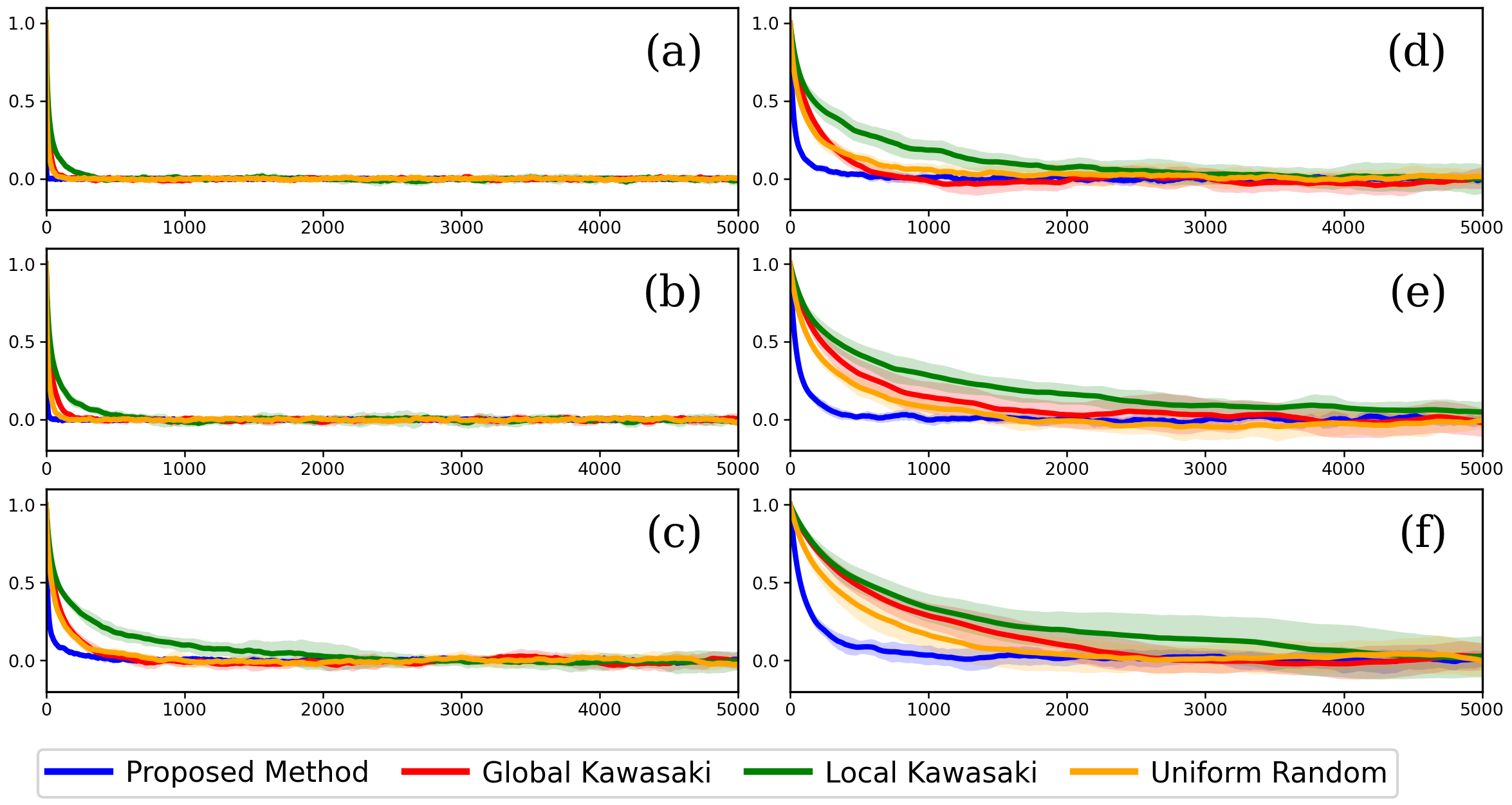}
  \caption{Autocorrelation $\rho(l)$ for 3-regular graphs. The block size is fixed as $|B| = 16$, and the system size $N$ is increased. The yellow curves denote uniform block resampling using the same partitions as the proposed method. The vertical axis shows the autocorrelation $\rho(l)$ of the overlap $q(t)$, and the horizontal axis represents the lag $l$. Faster decay indicates more rapid mixing. Panels show (a) $N=16$, (b) $N=32$, (c) $N=64$, (d) $N=128$, (e) $N=256$, (f) $N=512$.}
  \label{fig:3regular_step_syssize}
\end{figure}

\begin{figure}[t]
  \centering
  \includegraphics[width=1.0\linewidth]{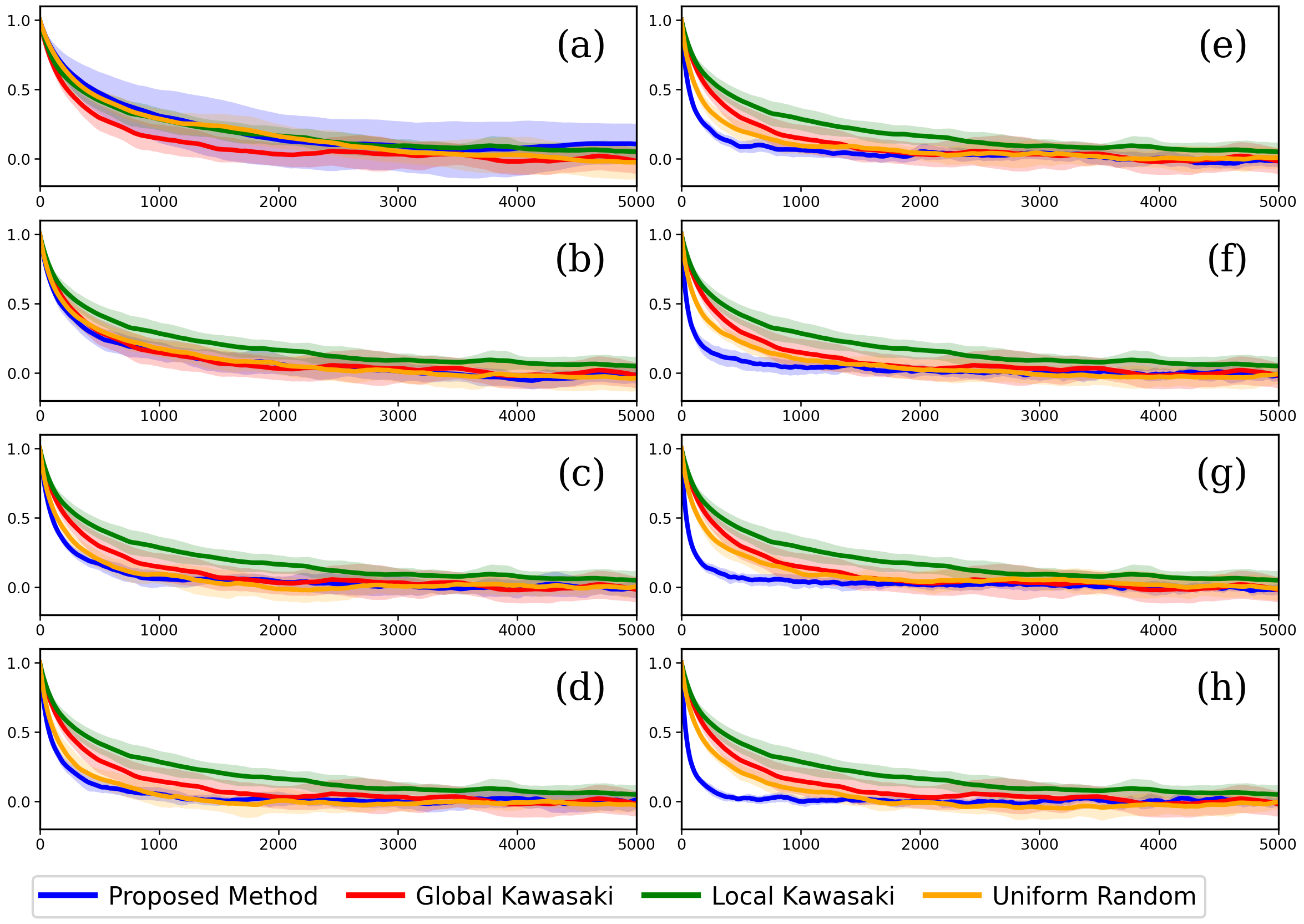}
  \caption{Autocorrelation $\rho(l)$ for a 3-regular graph. The system size is fixed as $N = 256$, and the nominal block size $|B|$ is increased. The yellow curves denote uniform block resampling using the same partitions as the proposed method. The vertical axis shows the autocorrelation $\rho(l)$ of the overlap $q(t)$, and the horizontal axis represents the lag $l$. Faster decay indicates more rapid mixing. Panels show (a) $|B|=2$, (b) $|B|=4$, (c) $|B|=6$, (d) $|B|=8$, (e) $|B|=10$, (f) $|B|=12$, (g) $|B|=14$, (h) $|B|=16$.}
  \label{fig:3regular_step_blocksize}
\end{figure}

Here we provide additional details for the experiments presented in Sec.~\ref{sec:3regular}.
The number of QAOA layers is set to $p=5$.
We use the target inverse temperature \mbox{$\beta_{\pi}=0.5$}, chosen to approximately match the representative fitted effective inverse temperatures \mbox{$\beta_{\mathrm{fit}}$} of the blockwise QAOA sample distributions.
To obtain the results shown in Fig.~\ref{fig:3regular_system} and Fig.~\ref{fig:3regular_block}, we first compute the raw data as a function of the MCMC steps.
These data are presented in Fig.~\ref{fig:3regular_step_syssize} and Fig.~\ref{fig:3regular_step_blocksize}, respectively.
We fit the decay using the exponential form \mbox{$\exp(-\tau l)$}, with the amplitude fixed by \mbox{$\rho(0)=1$}, and define the fitted parameter $\tau$ as the decay rate.

\begin{figure}[t]
  \centering
  \includegraphics[width=1.0\linewidth]{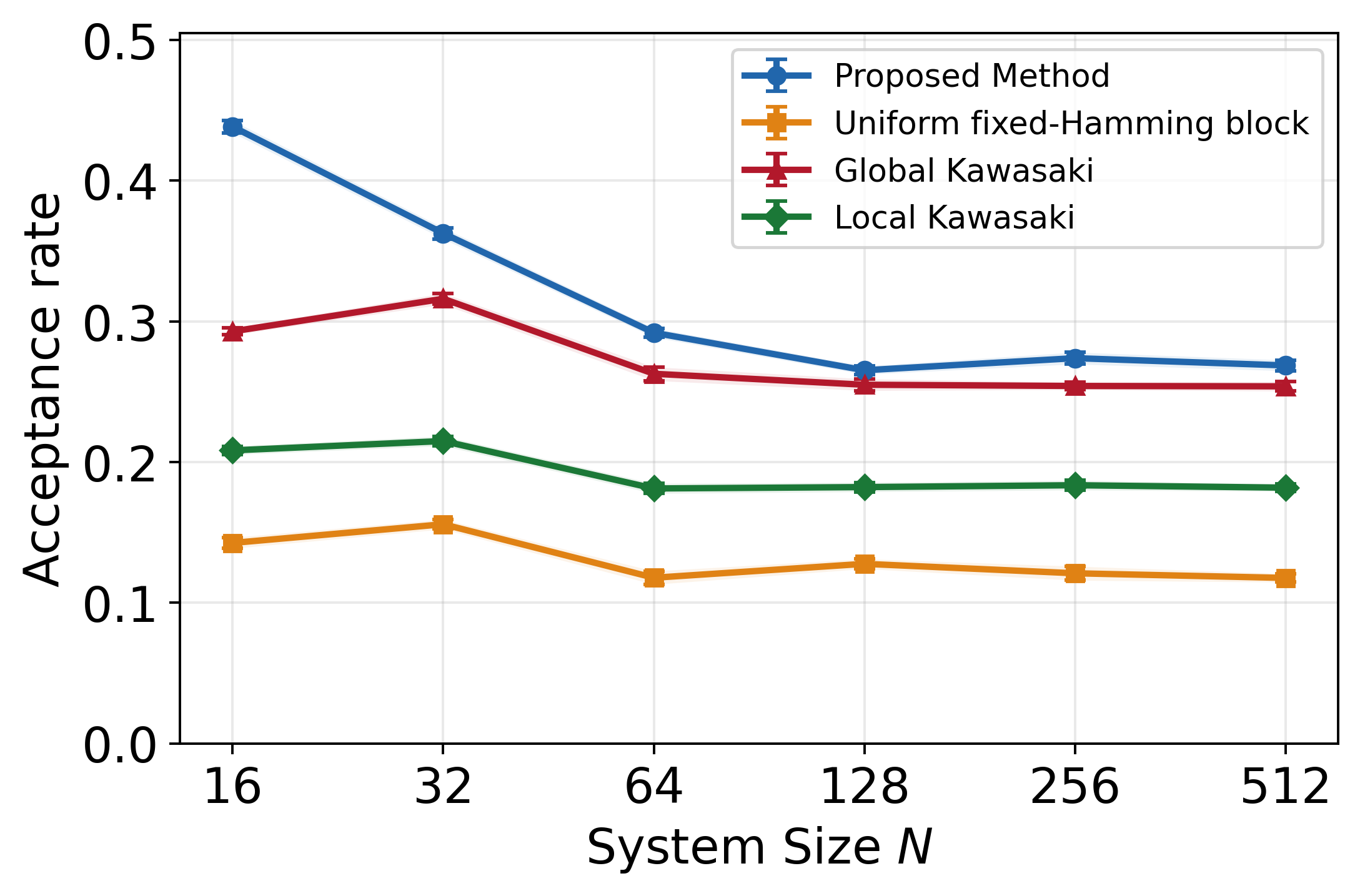}
  \caption{Acceptance rate as a function of the system size
  \mbox{$N$} for the 3-regular-graph experiments.
  The block size is fixed at \mbox{$|B|=16$}.
  The proposed QAOA--MADE block proposal is compared with uniform
  resampling within the selected block at the same fixed block Hamming weight.}
  \label{fig:3regular_system_acceptance}
\end{figure}

\begin{figure}[t]
  \centering
  \includegraphics[width=1.0\linewidth]{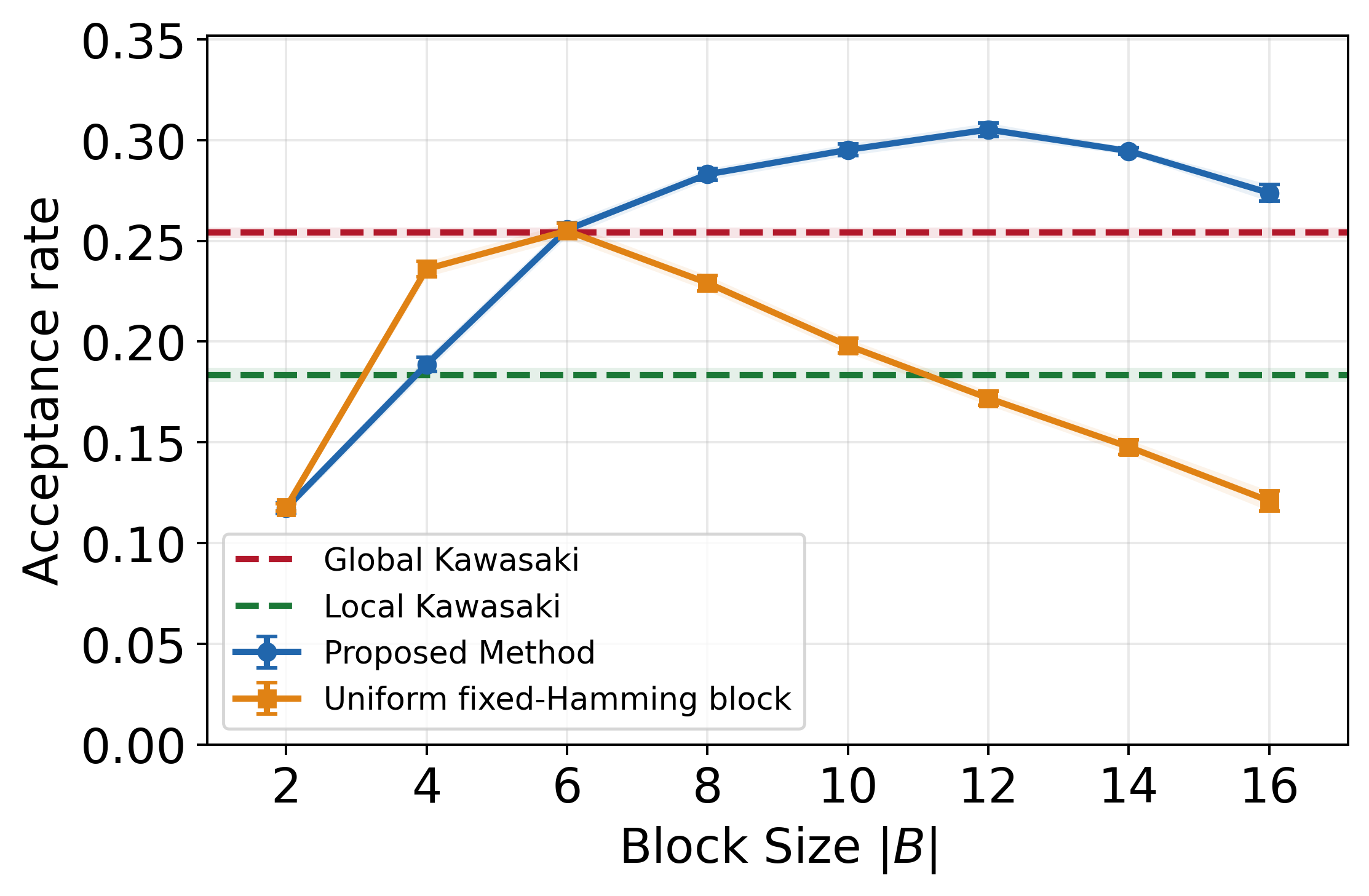}
  \caption{Acceptance rate as a function of the nominal block size
  \mbox{$|B|$} for the 3-regular-graph experiments with
  \mbox{$N=256$}.
  The proposed QAOA--MADE block proposal is compared with uniform
  resampling within the selected block at the same fixed block Hamming weight.}
  \label{fig:3regular_block_acceptance}
\end{figure}

Figures~\mbox{\ref{fig:3regular_system_acceptance}} and
\mbox{\ref{fig:3regular_block_acceptance}} show the acceptance rates for
the system-size and block-size sweeps, respectively.
As the block size increases, the acceptance rate of the uniform block proposal
decreases substantially, consistently with the deterioration of its fitted
decay rate \mbox{$\tau$}.
Because the two methods use the same partitions at each block size, the difference
in their block-size dependence cannot be explained by the partitioning alone.
By contrast, the QAOA--MADE proposal maintains a higher acceptance rate for larger
blocks, consistent with its capturing useful internal-energy structure and
producing proposals better adapted to constrained Boltzmann sampling.
For the system-size sweep at fixed \mbox{$|B|=16$}, the acceptance rates show no
systematic decrease with increasing \mbox{$N$}.
This is consistent with acceptance being governed primarily by the selected
block and its boundary interactions, so that a weak dependence on the total
system size is expected at fixed block size and comparable local graph statistics.
The nearly size-independent acceptance rates indicate that increasing
\mbox{$N$} does not introduce an additional rejection penalty for these block
proposals.
This is consistent with the persistence of the observed constant-factor
advantage, although it does not imply that the absolute decay rate
\mbox{$\tau$} is independent of \mbox{$N$}.

\begin{figure*}[t]
  \centering
  \includegraphics[width=1.0\linewidth]{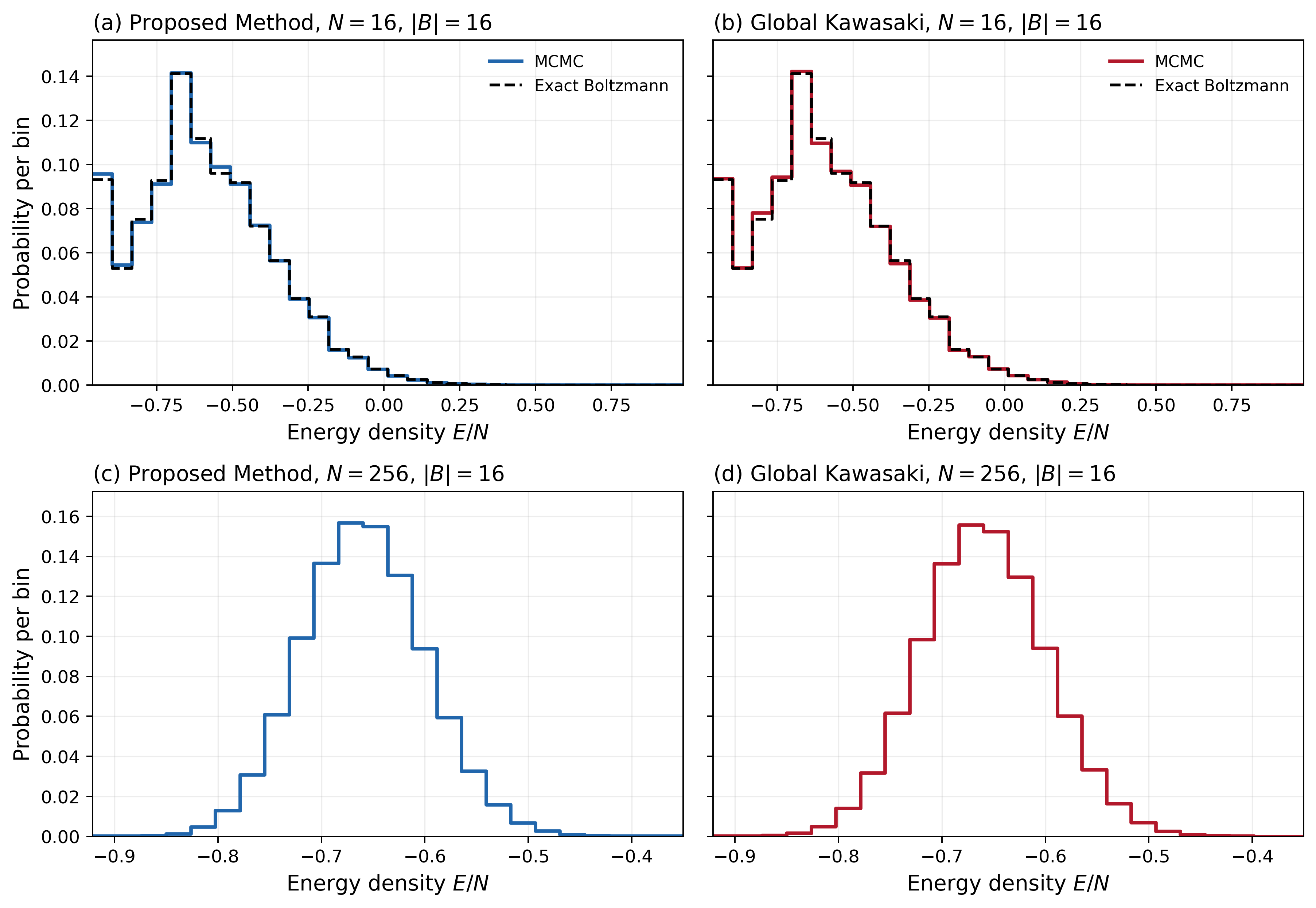}
    \caption{Normalized energy-density histograms for constrained Boltzmann sampling on 3-regular graphs at \mbox{$\beta_{\pi}=0.5$} and \mbox{$K=N/2$}. Panels (a) and (b) show the proposed method and global Kawasaki dynamics, respectively, for \mbox{$N=16$} and \mbox{$|B|=16$}; the black dashed lines show the exact constrained Boltzmann distribution. Panels (c) and (d) show the corresponding results for \mbox{$N=256$} and \mbox{$|B|=16$}. Each histogram is obtained from a single post-burn-in MCMC trajectory, and common bin edges are used within each row.}
    \label{fig:3regular_energy_histogram}
\end{figure*}

To verify that the faster decorrelation does not come at the cost of distorting the target energy distribution, Fig.~\mbox{\ref{fig:3regular_energy_histogram}} compares normalized histograms of the energy density \mbox{$\varepsilon=E/N$} obtained using the proposed method and global Kawasaki dynamics at \mbox{$\beta_{\pi}=0.5$}, \mbox{$K=N/2$}, and \mbox{$|B|=16$}.
Each panel was constructed from a single long MCMC trajectory without thinning. For \mbox{$N=16$}, the first \mbox{$5.0\times10^{4}$} transitions were discarded as burn-in and the subsequent \mbox{$5.0\times10^{5}$} states were retained, whereas for \mbox{$N=256$} the corresponding numbers were \mbox{$2.0\times10^{5}$} and \mbox{$5.0\times10^{6}$}, respectively.
A common set of 30 bins for \mbox{$N=16$} and 24 bins for \mbox{$N=256$} was used for the two sampling methods.
For \mbox{$N=16$}, we additionally enumerated all \mbox{$\binom{16}{8}=12{,}870$} feasible configurations and evaluated their Boltzmann weights exactly; the resulting exact constrained Boltzmann distribution \mbox{$\pi_K$} is shown by the black dashed lines in panels (a) and (b).
We quantify the agreement primarily using the one-dimensional 1-Wasserstein distance
\begin{align}
W_1(P,Q)
=
\int_{-\infty}^{\infty}
\left|F_P(\varepsilon)-F_Q(\varepsilon)\right|
\,d\varepsilon ,
\end{align}
where \mbox{$F_P$} and \mbox{$F_Q$} are the cumulative distribution functions of \mbox{$\varepsilon$}.
This metric is independent of the histogram bins and measures the displacement between two distributions along the physically ordered energy axis. For \mbox{$N=16$}, we obtain \mbox{$W_1=1.50\times10^{-3}$} between the proposed method and the exact distribution, \mbox{$2.11\times10^{-3}$} between global Kawasaki dynamics and the exact distribution, and \mbox{$1.96\times10^{-3}$} between the two samplers.
For \mbox{$N=256$}, the distance between the proposed method and global Kawasaki dynamics is \mbox{$5.39\times10^{-4}$}.
All these distances are less than \mbox{$1\%$} of the standard deviation of the corresponding energy distributions. As a complementary measure tied directly to the displayed histograms, we also calculate the binned total-variation distance, \mbox{$D_{\mathrm{TV}}^{(\mathrm{bin})}=\frac{1}{2}\sum_b|p_b-q_b|$}.
For the same four comparisons, the values are \mbox{$7.59\times10^{-3}$}, \mbox{$6.79\times10^{-3}$}, \mbox{$9.14\times10^{-3}$}, and \mbox{$5.96\times10^{-3}$}, respectively, all below \mbox{$10^{-2}$}.
Thus, the sampled energy distributions show excellent agreement with the exact result when it is available and across the two sampling schemes for the larger system.
Agreement of this one-dimensional energy marginal alone does not constitute a proof of irreducibility or ergodicity, and cannot exclude incomplete exploration among configurations with identical or nearly identical energies.
Nevertheless, together with the autocorrelation analysis, it provides a stringent empirical consistency check and substantially strengthens confidence in the reliability of the sampled distributions.

\begin{figure}[t]
  \centering
  \includegraphics[width=0.8\linewidth]{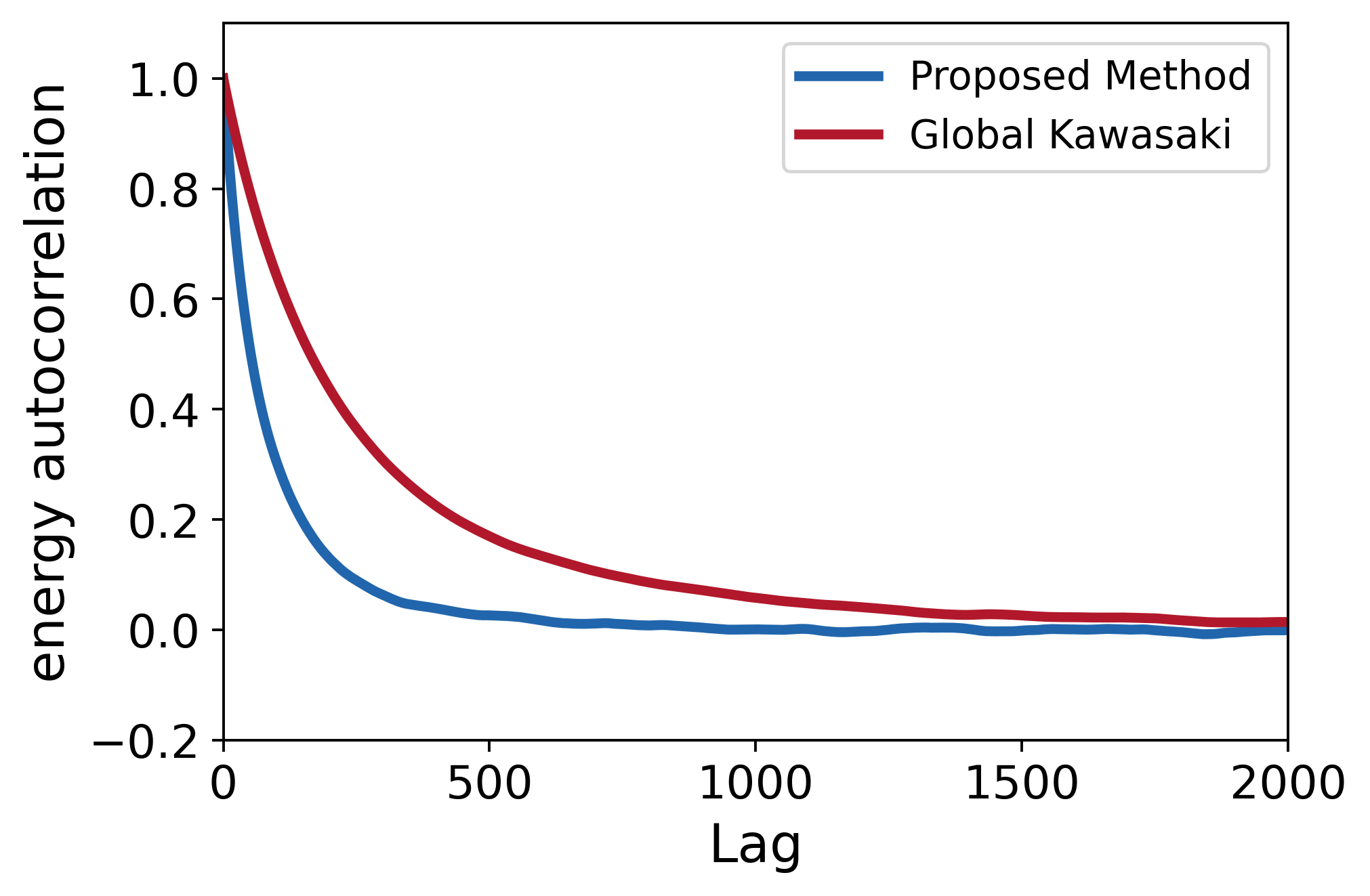}
  \caption{Energy autocorrelation for constrained Boltzmann sampling on a 3-regular graph with \mbox{$N=256$}, \mbox{$|B|=16$}, \mbox{$\beta_{\pi}=0.5$}, and \mbox{$K=N/2$}. The blue and red curves denote the proposed method and global Kawasaki dynamics, respectively. Each curve is calculated directly from a single MCMC trajectory after discarding the first \mbox{$2.0\times10^{5}$} transitions as burn-in and retaining the subsequent \mbox{$5.0\times10^{6}$} states without thinning. The vertical axis shows the autocorrelation of the energy density \mbox{$E/N$}, and the horizontal axis denotes the lag. Faster decay indicates more rapid decorrelation of the sampled energies.}
  \label{fig:3regular_energy_autocorrelation}
\end{figure}

To complement the energy-distribution comparison above,
we examine the autocorrelation of the sampled energy density,
\mbox{$\varepsilon(t)=E(t)/N$}, along the same post-burn-in trajectories.
As shown in Fig.~\mbox{\ref{fig:3regular_energy_autocorrelation}},
the energy autocorrelation decays more rapidly for the proposed
method than for global Kawasaki dynamics.
Together with the agreement of the energy distributions, this
provides complementary empirical support for the reliability
and efficiency of the proposed method under the tested conditions.
These observable-level checks do not by themselves establish
complete exploration of the constrained configuration space.

The target inverse temperature \mbox{$\beta_{\pi}=0.5$} used in the
3-regular-graph experiments was chosen as a single representative value
that best characterizes the effective inverse temperatures of the
post-QAOA distributions across the blocks.  Because both the QAOA
distribution and the block Hamiltonian vary from block to block, we first
estimated an effective inverse temperature \mbox{$\beta_{\mathrm{fit}}$} for each
block and then selected \mbox{$\beta_{\pi}=0.5$} as a common value near the center
of these blockwise estimates.  Using one common value, rather than
retuning the target temperature for each block or numerical setting, also
keeps the comparison between the proposed method and Kawasaki dynamics
under the same sampling condition.  Thus, this procedure empirically
calibrates the temperature scale relevant to the QAOA proposals; it does
not assume that \mbox{$\beta=0.5$} is a universal thermodynamic low-temperature
regime for random 3-regular graphs.

\begin{figure*}[t]
  \centering
  \includegraphics[width=\textwidth]{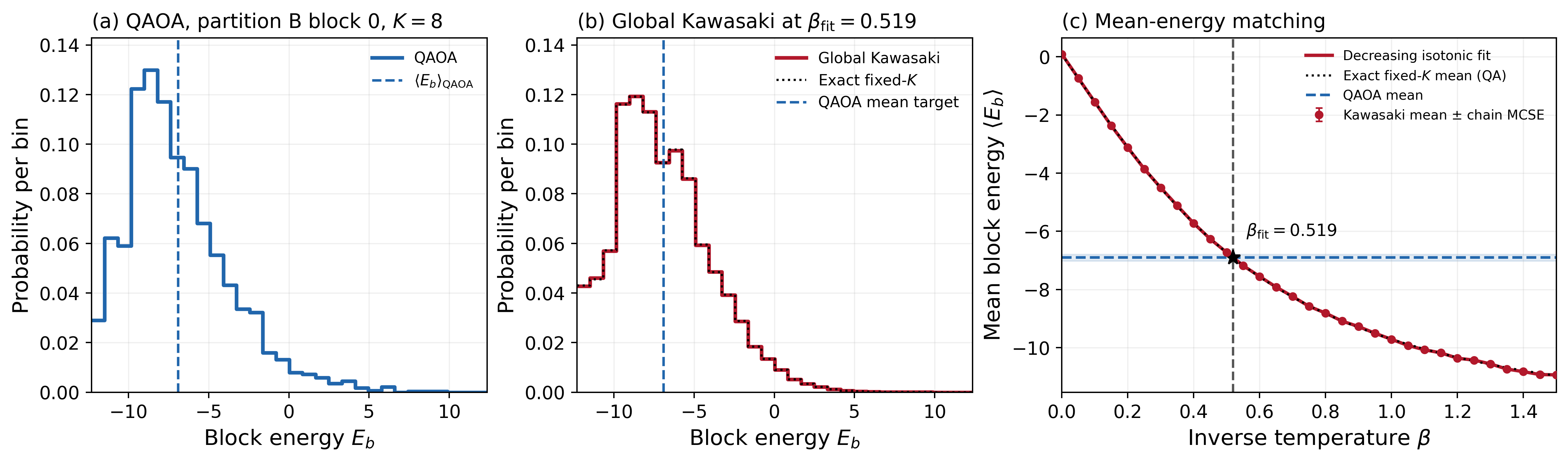}
  \caption{Construction of the blockwise effective inverse temperature
  for a representative block with \mbox{$N=256$}, \mbox{$|B|=16$}, and \mbox{$K_m=8$}
  (partition B, block 0).
  (a) Energy histogram of the QAOA output conditioned on \mbox{$K_m=8$}; the
  dashed vertical line denotes its mean block energy.
  (b) Energy histogram obtained using global Kawasaki dynamics at the
  fitted inverse temperature, together with the exact fixed-\mbox{$K_m$}
  Boltzmann histogram.  Their agreement verifies that the block-level
  Kawasaki chain reproduces its equilibrium energy distribution.
  (c) Mean block energy obtained using global Kawasaki dynamics as a
  function of \mbox{$\beta$}.  The points and error bars show the Markov-chain
  mean and its Monte Carlo standard error, the solid curve is a decreasing
  isotonic fit, and the dotted curve is the exact fixed-\mbox{$K_m$} result.  The
  intersection with the QAOA mean defines \mbox{$\beta_{\mathrm{fit}}$} and gives
  \mbox{$\beta_{\mathrm{fit}}=0.519$} for this block.}
  \label{fig:beta_fit_calibration}
\end{figure*}

The fitting procedure is illustrated in
Fig.~\mbox{\ref{fig:beta_fit_calibration}}.  For each block \mbox{$B_m$}, we consider
the central Hamming-weight sector
\mbox{$K_m=\lfloor |B_m|/2\rfloor$}, which is representative of the global
half-filling constraint, and calculate the mean internal block energy of
the QAOA output conditioned on this sector.  We then perform global
Kawasaki sampling for the same block and Hamming weight over a grid of
trial inverse temperatures.  A monotonically decreasing curve is fitted
to the resulting Kawasaki mean energies, and the blockwise effective
inverse temperature is defined by the mean-energy matching condition
\begin{equation}
  \left\langle E_{B_m}\right\rangle_
  {\mathrm{Kawasaki};\,K_m,\,\beta=\beta_{\mathrm{fit}}^{(m)}}
  =
  \left\langle E_{B_m}\right\rangle_{\mathrm{QAOA};\,K_m}.
  \label{eq:beta_fit_definition}
\end{equation}
In other words, \mbox{$\beta_{\mathrm{fit}}^{(m)}$} is the inverse temperature at
which Kawasaki dynamics reproduces the post-QAOA mean block energy.  The
representative energy histograms in
Figs.~\mbox{\ref{fig:beta_fit_calibration}}(a) and
\mbox{\ref{fig:beta_fit_calibration}}(b) also show that the associated
distributions have similar shapes, although the definition of
\mbox{$\beta_{\mathrm{fit}}$} itself uses only their first moments.

Table~\mbox{\ref{tab:beta_fit_summary}} summarizes the blockwise fits for all
\mbox{$(N,|B|)$} settings used in the 3-regular-graph experiments.  The reported
standard deviation describes heterogeneity among the blocks and is not a
standard error of the setting-level mean.  For this full scan, the
equilibrium Kawasaki mean was evaluated by exact enumeration within each
fixed-Hamming-weight sector, which is the infinite-chain limit of the
MCMC calibration shown in Fig.~\mbox{\ref{fig:beta_fit_calibration}} and removes
finite-chain noise from the setting-to-setting comparison.  For \mbox{$|B|=2$},
the internal energy is constant in the \mbox{$K_m=1$} sector, so an inverse
temperature cannot be identified.  For \mbox{$|B|=4$}, the mean-energy curve can
become nearly flat because it contains only a small number of energy
levels; its inversion is therefore ill-conditioned and produces the
large mean and standard deviation shown in the table.  The median of all
12 finite setting-level means is \mbox{$0.5070$}, and the unweighted mean over
the 11 well-resolved settings with nominal \mbox{$|B|\geq 6$} is \mbox{$0.5129$}.
Therefore, the typical fitted inverse temperature is close to \mbox{$0.5$}, and
we fixed \mbox{$\beta_{\pi}=0.5$} in all 3-regular-graph experiments to approximately match the
QAOA effective temperature while maintaining identical conditions across
system and block sizes.

\begin{table}[t]
  \centering
  \caption{Mean and sample standard deviation of the blockwise
  \mbox{$\beta_{\mathrm{fit}}$} values over both block partitions for each
  numerical setting.  Only blocks for which the inverse temperature is
  identifiable are included.  The unweighted mean of the 11
  well-resolved setting-level means with nominal \mbox{$|B|\geq6$} is \mbox{$0.5129$}.
  For completeness, the mean over all 12 finite rows is \mbox{$0.6083$} and
  their median is \mbox{$0.5070$}; the former is dominated by the
  ill-conditioned \mbox{$|B|=4$} row.  A dash indicates that
  \mbox{$\beta_{\mathrm{fit}}$} is not identifiable.}
  \label{tab:beta_fit_summary}
  \renewcommand{\arraystretch}{1.12}
  \begin{ruledtabular}
    \begin{tabular}{cccc}
      \mbox{$N$} & \mbox{$|B|$} & \mbox{$\overline{\beta}_{\mathrm{fit}}$}
      & \mbox{$\mathrm{SD}(\beta_{\mathrm{fit}})$} \\
      \hline
       16 & 16 & 0.4372 & 0.0339 \\
       32 & 16 & 0.5140 & 0.0514 \\
       64 & 16 & 0.4712 & 0.0435 \\
      128 & 16 & 0.4999 & 0.2681 \\
      256 & 16 & 0.4784 & 0.1790 \\
      512 & 16 & 0.4784 & 0.1387 \\
      256 &  2 & --     & --     \\
      256 &  4 & 1.6582 & 2.7226 \\
      256 &  6 & 0.6618 & 0.3977 \\
      256 &  8 & 0.5518 & 0.2570 \\
      256 & 10 & 0.5573 & 0.2421 \\
      256 & 12 & 0.5205 & 0.1963 \\
      256 & 14 & 0.4711 & 0.1330 \\
    \end{tabular}
  \end{ruledtabular}
\end{table}

\section{MNIST Feature Selection}\label{ap:mnist}
We provide additional details for the experiments presented in Sec.~\ref{sec:mnist}.
We first define the objective function.
Let the training dataset be $\mathcal{D}=\{(\bm{z}^{(n)},y^{(n)})\}_{n=1}^{N_{\mathrm{data}}}$, where
$\bm{z}^{(n)}\in\mathbb{R}^{N}$ is the vectorized image of sample $n$ and $y^{(n)}\in\mathcal{C}$ is its label.
$\mathcal{C}$ denotes the set of class labels.
We introduce binary selection variables $\bm{x}\in\{0,1\}^{N}$, where $x_i=1$ indicates that pixel $i$ is selected in the mask.
We quantify the dependence between a pixel value and the label by the mutual information computed from the training data.
Assuming binarized pixels $z_i\in\{0,1\}$, we define
\begin{align}
I&(z_i;y):=  \notag \\
&\sum_{v\in\{0,1\}}\sum_{c\in\mathcal{C}}
p(z_i=v,y=c)\log\frac{p(z_i=v,y=c)}{p(z_i=v)\,p(y=c)},
\label{eq:mi_def_zi_y}
\end{align}
where $p(z_i=v,y=c)$ is the joint probability of observing pixel value $v$ at pixel $i$ and label $c$,
$p(z_i=v)$ is the marginal probability of observing pixel value $v$ at pixel $i$,
and $p(y=c)$ is the marginal probability of observing label $c$.
All probabilities are estimated from the training dataset $\mathcal{D}$ via empirical frequencies.
We compute $I(z_i;z_j)$ analogously from the same training dataset.
We then construct the energy:
\begin{align}
E(\bm{x}) = &- \sum_{i=1}^{N} I(z_i; y)\, x_i  \notag \\
&+ \frac{1}{K-1}\sum_{1\le i<j\le N} I(z_i; z_j)\, x_i x_j ,
\end{align}
where the factor $(K-1)^{-1}$ is included for normalization.
Intuitively, $I(z_i; y)$ measures the statistical dependence between the pixels and the labels.
Stronger dependence lowers the energy, meaning that the corresponding pixels are favored and are expected to be useful for classification.
In addition, $I(z_i; z_j)$ captures the dependence between two pixels: when two pixels are less redundant, selecting them together tends to lower the energy, so the optimization encourages choosing diverse pixels.
By optimizing this energy under the fixed Hamming weight constraint $\sum_{i=1}^{N} x_i = K$, the $K$ pixels whose indices satisfy $x_i=1$ form the mask.
We then perform training and inference using only these selected pixels with a logistic regression classifier.
The corresponding interaction graph tends to be sparse because pixels that are far apart usually have weak correlations.
Therefore, we remove connections whose coefficients are below a threshold.

\begin{figure}[t]
  \centering
  \includegraphics[width=0.9\linewidth]{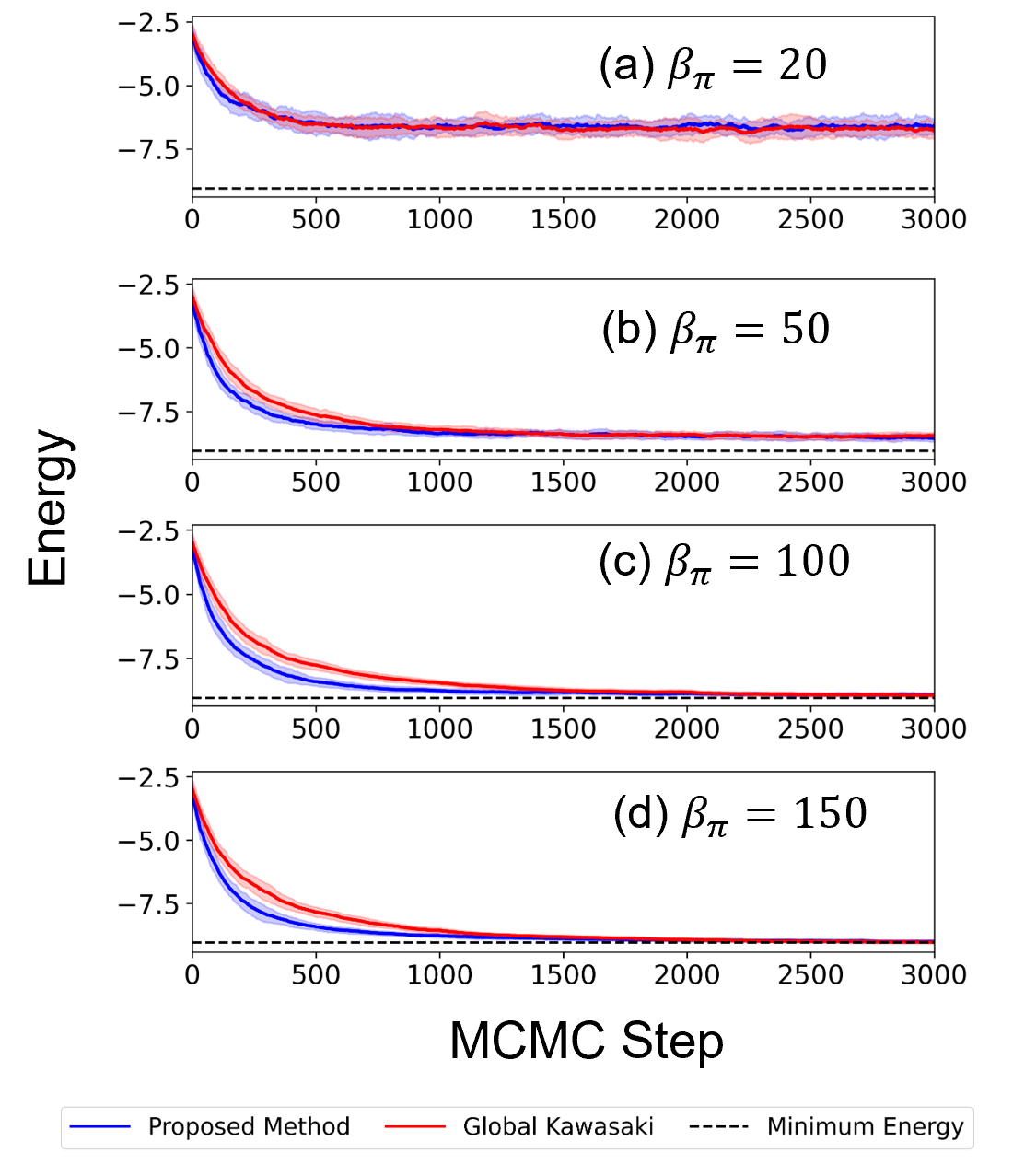}
  \caption{MCMC results for MNIST feature selection. The horizontal axis shows the MCMC step, and the vertical axis shows the energy. The blue curve denotes the proposed method, the red curve denotes global Kawasaki dynamics, and the black dashed line denotes the best-found energy obtained by global Kawasaki dynamics across the tested runs and inverse temperatures within the MCMC step budgets used here. This reference value is not a certified constrained ground-state energy. The solid curves show the mean energy across replicas, and the
   shaded regions indicate \mbox{$\pm1$} standard deviation across replicas at each step. (a) $\beta_{\pi} = 20$, (b) $\beta_{\pi} = 50$, (c) $\beta_{\pi} = 100$, and (d) $\beta_{\pi} = 150$.}
  \label{fig:mnist_appendix}
\end{figure}

We provide additional results.
Figure~\ref{fig:mnist_appendix} shows the results of MCMC runs with different values of the parameter $\beta_{\pi}$.
Note that the vertical axis represents the energy itself, not the best energy.
When $\beta_{\pi}$ is small, the output from MADE is biased strongly toward the low-temperature distribution, which reduces the acceptance rate.
As a result, the energy converges to a level similar to that of the Kawasaki method.
As $\beta_{\pi}$ increases, our method becomes more advantageous.
In particular, around $\beta_{\pi} = 100$, the shape of the MCMC convergence changes very little beyond this point, and the method can still reach sufficiently low energies.
For this reason, we used $\beta_{\pi} = 100$ in the comparison of classification performance in Sec.~\ref{sec:mnist}.

\bibliographystyle{apsrev4-2}

\begin{thebibliography}{41}%
\makeatletter
\providecommand \@ifxundefined [1]{%
 \@ifx{#1\undefined}
}%
\providecommand \@ifnum [1]{%
 \ifnum #1\expandafter \@firstoftwo
 \else \expandafter \@secondoftwo
 \fi
}%
\providecommand \@ifx [1]{%
 \ifx #1\expandafter \@firstoftwo
 \else \expandafter \@secondoftwo
 \fi
}%
\providecommand \natexlab [1]{#1}%
\providecommand \enquote  [1]{``#1''}%
\providecommand \bibnamefont  [1]{#1}%
\providecommand \bibfnamefont [1]{#1}%
\providecommand \citenamefont [1]{#1}%
\providecommand \href@noop [0]{\@secondoftwo}%
\providecommand \href [0]{\begingroup \@sanitize@url \@href}%
\providecommand \@href[1]{\@@startlink{#1}\@@href}%
\providecommand \@@href[1]{\endgroup#1\@@endlink}%
\providecommand \@sanitize@url [0]{\catcode `\\12\catcode `\$12\catcode `\&12\catcode `\#12\catcode `\^12\catcode `\_12\catcode `\%12\relax}%
\providecommand \@@startlink[1]{}%
\providecommand \@@endlink[0]{}%
\providecommand \url  [0]{\begingroup\@sanitize@url \@url }%
\providecommand \@url [1]{\endgroup\@href {#1}{\urlprefix }}%
\providecommand \urlprefix  [0]{URL }%
\providecommand \Eprint [0]{\href }%
\providecommand \doibase [0]{https://doi.org/}%
\providecommand \selectlanguage [0]{\@gobble}%
\providecommand \bibinfo  [0]{\@secondoftwo}%
\providecommand \bibfield  [0]{\@secondoftwo}%
\providecommand \translation [1]{[#1]}%
\providecommand \BibitemOpen [0]{}%
\providecommand \bibitemStop [0]{}%
\providecommand \bibitemNoStop [0]{.\EOS\space}%
\providecommand \EOS [0]{\spacefactor3000\relax}%
\providecommand \BibitemShut  [1]{\csname bibitem#1\endcsname}%
\let\auto@bib@innerbib\@empty
\bibitem [{\citenamefont {Shor}(1994)}]{shor1994algorithms}%
  \BibitemOpen
  \bibfield  {author} {\bibinfo {author} {\bibfnamefont {P.~W.}\ \bibnamefont {Shor}},\ }in\ \href@noop {} {\emph {\bibinfo {booktitle} {Proceedings 35th annual symposium on foundations of computer science}}}\ (\bibinfo {organization} {Ieee},\ \bibinfo {year} {1994})\ pp.\ \bibinfo {pages} {124--134}\BibitemShut {NoStop}%
\bibitem [{\citenamefont {Grover}(1997)}]{grover1997quantum}%
  \BibitemOpen
  \bibfield  {author} {\bibinfo {author} {\bibfnamefont {L.~K.}\ \bibnamefont {Grover}},\ }\href@noop {} {\bibfield  {journal} {\bibinfo  {journal} {Physical review letters}\ }\textbf {\bibinfo {volume} {79}},\ \bibinfo {pages} {325} (\bibinfo {year} {1997})}\BibitemShut {NoStop}%
\bibitem [{\citenamefont {Abrams}\ and\ \citenamefont {Lloyd}(1999)}]{abrams1999quantum}%
  \BibitemOpen
  \bibfield  {author} {\bibinfo {author} {\bibfnamefont {D.~S.}\ \bibnamefont {Abrams}}\ and\ \bibinfo {author} {\bibfnamefont {S.}~\bibnamefont {Lloyd}},\ }\href@noop {} {\bibfield  {journal} {\bibinfo  {journal} {Physical Review Letters}\ }\textbf {\bibinfo {volume} {83}},\ \bibinfo {pages} {5162} (\bibinfo {year} {1999})}\BibitemShut {NoStop}%
\bibitem [{\citenamefont {Harrow}\ \emph {et~al.}(2009)\citenamefont {Harrow}, \citenamefont {Hassidim},\ and\ \citenamefont {Lloyd}}]{harrow2009quantum}%
  \BibitemOpen
  \bibfield  {author} {\bibinfo {author} {\bibfnamefont {A.~W.}\ \bibnamefont {Harrow}}, \bibinfo {author} {\bibfnamefont {A.}~\bibnamefont {Hassidim}},\ and\ \bibinfo {author} {\bibfnamefont {S.}~\bibnamefont {Lloyd}},\ }\href@noop {} {\bibfield  {journal} {\bibinfo  {journal} {Physical review letters}\ }\textbf {\bibinfo {volume} {103}},\ \bibinfo {pages} {150502} (\bibinfo {year} {2009})}\BibitemShut {NoStop}%
\bibitem [{\citenamefont {Preskill}(2018)}]{preskill2018quantum}%
  \BibitemOpen
  \bibfield  {author} {\bibinfo {author} {\bibfnamefont {J.}~\bibnamefont {Preskill}},\ }\href@noop {} {\bibfield  {journal} {\bibinfo  {journal} {Quantum}\ }\textbf {\bibinfo {volume} {2}},\ \bibinfo {pages} {79} (\bibinfo {year} {2018})}\BibitemShut {NoStop}%
\bibitem [{\citenamefont {Arute}\ \emph {et~al.}(2019)\citenamefont {Arute}, \citenamefont {Arya}, \citenamefont {Babbush}, \citenamefont {Bacon}, \citenamefont {Bardin}, \citenamefont {Barends}, \citenamefont {Biswas}, \citenamefont {Boixo}, \citenamefont {Brandao}, \citenamefont {Buell} \emph {et~al.}}]{arute2019quantum}%
  \BibitemOpen
  \bibfield  {author} {\bibinfo {author} {\bibfnamefont {F.}~\bibnamefont {Arute}}, \bibinfo {author} {\bibfnamefont {K.}~\bibnamefont {Arya}}, \bibinfo {author} {\bibfnamefont {R.}~\bibnamefont {Babbush}}, \bibinfo {author} {\bibfnamefont {D.}~\bibnamefont {Bacon}}, \bibinfo {author} {\bibfnamefont {J.~C.}\ \bibnamefont {Bardin}}, \bibinfo {author} {\bibfnamefont {R.}~\bibnamefont {Barends}}, \bibinfo {author} {\bibfnamefont {R.}~\bibnamefont {Biswas}}, \bibinfo {author} {\bibfnamefont {S.}~\bibnamefont {Boixo}}, \bibinfo {author} {\bibfnamefont {F.~G.}\ \bibnamefont {Brandao}}, \bibinfo {author} {\bibfnamefont {D.~A.}\ \bibnamefont {Buell}}, \emph {et~al.},\ }\href@noop {} {\bibfield  {journal} {\bibinfo  {journal} {nature}\ }\textbf {\bibinfo {volume} {574}},\ \bibinfo {pages} {505} (\bibinfo {year} {2019})}\BibitemShut {NoStop}%
\bibitem [{\citenamefont {Zhong}\ \emph {et~al.}(2020)\citenamefont {Zhong}, \citenamefont {Wang}, \citenamefont {Deng}, \citenamefont {Chen}, \citenamefont {Peng}, \citenamefont {Luo}, \citenamefont {Qin}, \citenamefont {Wu}, \citenamefont {Ding}, \citenamefont {Hu} \emph {et~al.}}]{zhong2020quantum}%
  \BibitemOpen
  \bibfield  {author} {\bibinfo {author} {\bibfnamefont {H.-S.}\ \bibnamefont {Zhong}}, \bibinfo {author} {\bibfnamefont {H.}~\bibnamefont {Wang}}, \bibinfo {author} {\bibfnamefont {Y.-H.}\ \bibnamefont {Deng}}, \bibinfo {author} {\bibfnamefont {M.-C.}\ \bibnamefont {Chen}}, \bibinfo {author} {\bibfnamefont {L.-C.}\ \bibnamefont {Peng}}, \bibinfo {author} {\bibfnamefont {Y.-H.}\ \bibnamefont {Luo}}, \bibinfo {author} {\bibfnamefont {J.}~\bibnamefont {Qin}}, \bibinfo {author} {\bibfnamefont {D.}~\bibnamefont {Wu}}, \bibinfo {author} {\bibfnamefont {X.}~\bibnamefont {Ding}}, \bibinfo {author} {\bibfnamefont {Y.}~\bibnamefont {Hu}}, \emph {et~al.},\ }\href@noop {} {\bibfield  {journal} {\bibinfo  {journal} {Science}\ }\textbf {\bibinfo {volume} {370}},\ \bibinfo {pages} {1460} (\bibinfo {year} {2020})}\BibitemShut {NoStop}%
\bibitem [{\citenamefont {Madsen}\ \emph {et~al.}(2022)\citenamefont {Madsen}, \citenamefont {Laudenbach}, \citenamefont {Askarani}, \citenamefont {Rortais}, \citenamefont {Vincent}, \citenamefont {Bulmer}, \citenamefont {Miatto}, \citenamefont {Neuhaus}, \citenamefont {Helt}, \citenamefont {Collins} \emph {et~al.}}]{madsen2022quantum}%
  \BibitemOpen
  \bibfield  {author} {\bibinfo {author} {\bibfnamefont {L.~S.}\ \bibnamefont {Madsen}}, \bibinfo {author} {\bibfnamefont {F.}~\bibnamefont {Laudenbach}}, \bibinfo {author} {\bibfnamefont {M.~F.}\ \bibnamefont {Askarani}}, \bibinfo {author} {\bibfnamefont {F.}~\bibnamefont {Rortais}}, \bibinfo {author} {\bibfnamefont {T.}~\bibnamefont {Vincent}}, \bibinfo {author} {\bibfnamefont {J.~F.}\ \bibnamefont {Bulmer}}, \bibinfo {author} {\bibfnamefont {F.~M.}\ \bibnamefont {Miatto}}, \bibinfo {author} {\bibfnamefont {L.}~\bibnamefont {Neuhaus}}, \bibinfo {author} {\bibfnamefont {L.~G.}\ \bibnamefont {Helt}}, \bibinfo {author} {\bibfnamefont {M.~J.}\ \bibnamefont {Collins}}, \emph {et~al.},\ }\href@noop {} {\bibfield  {journal} {\bibinfo  {journal} {Nature}\ }\textbf {\bibinfo {volume} {606}},\ \bibinfo {pages} {75} (\bibinfo {year} {2022})}\BibitemShut {NoStop}%
\bibitem [{\citenamefont {Cerezo}\ \emph {et~al.}(2021)\citenamefont {Cerezo}, \citenamefont {Arrasmith}, \citenamefont {Babbush}, \citenamefont {Benjamin}, \citenamefont {Endo}, \citenamefont {Fujii}, \citenamefont {McClean}, \citenamefont {Mitarai}, \citenamefont {Yuan}, \citenamefont {Cincio} \emph {et~al.}}]{cerezo2021variational}%
  \BibitemOpen
  \bibfield  {author} {\bibinfo {author} {\bibfnamefont {M.}~\bibnamefont {Cerezo}}, \bibinfo {author} {\bibfnamefont {A.}~\bibnamefont {Arrasmith}}, \bibinfo {author} {\bibfnamefont {R.}~\bibnamefont {Babbush}}, \bibinfo {author} {\bibfnamefont {S.~C.}\ \bibnamefont {Benjamin}}, \bibinfo {author} {\bibfnamefont {S.}~\bibnamefont {Endo}}, \bibinfo {author} {\bibfnamefont {K.}~\bibnamefont {Fujii}}, \bibinfo {author} {\bibfnamefont {J.~R.}\ \bibnamefont {McClean}}, \bibinfo {author} {\bibfnamefont {K.}~\bibnamefont {Mitarai}}, \bibinfo {author} {\bibfnamefont {X.}~\bibnamefont {Yuan}}, \bibinfo {author} {\bibfnamefont {L.}~\bibnamefont {Cincio}}, \emph {et~al.},\ }\href@noop {} {\bibfield  {journal} {\bibinfo  {journal} {Nature Reviews Physics}\ }\textbf {\bibinfo {volume} {3}},\ \bibinfo {pages} {625} (\bibinfo {year} {2021})}\BibitemShut {NoStop}%
\bibitem [{\citenamefont {Farhi}\ \emph {et~al.}(2014)\citenamefont {Farhi}, \citenamefont {Goldstone},\ and\ \citenamefont {Gutmann}}]{farhi2014quantum}%
  \BibitemOpen
  \bibfield  {author} {\bibinfo {author} {\bibfnamefont {E.}~\bibnamefont {Farhi}}, \bibinfo {author} {\bibfnamefont {J.}~\bibnamefont {Goldstone}},\ and\ \bibinfo {author} {\bibfnamefont {S.}~\bibnamefont {Gutmann}},\ }\href@noop {} {\bibfield  {journal} {\bibinfo  {journal} {arXiv preprint arXiv:1411.4028}\ } (\bibinfo {year} {2014})}\BibitemShut {NoStop}%
\bibitem [{\citenamefont {Mitarai}\ \emph {et~al.}(2018)\citenamefont {Mitarai}, \citenamefont {Negoro}, \citenamefont {Kitagawa},\ and\ \citenamefont {Fujii}}]{mitarai2018quantum}%
  \BibitemOpen
  \bibfield  {author} {\bibinfo {author} {\bibfnamefont {K.}~\bibnamefont {Mitarai}}, \bibinfo {author} {\bibfnamefont {M.}~\bibnamefont {Negoro}}, \bibinfo {author} {\bibfnamefont {M.}~\bibnamefont {Kitagawa}},\ and\ \bibinfo {author} {\bibfnamefont {K.}~\bibnamefont {Fujii}},\ }\href@noop {} {\bibfield  {journal} {\bibinfo  {journal} {Physical Review A}\ }\textbf {\bibinfo {volume} {98}},\ \bibinfo {pages} {032309} (\bibinfo {year} {2018})}\BibitemShut {NoStop}%
\bibitem [{\citenamefont {Farhi}\ and\ \citenamefont {Neven}(2018)}]{farhi2018classification}%
  \BibitemOpen
  \bibfield  {author} {\bibinfo {author} {\bibfnamefont {E.}~\bibnamefont {Farhi}}\ and\ \bibinfo {author} {\bibfnamefont {H.}~\bibnamefont {Neven}},\ }\href@noop {} {\bibfield  {journal} {\bibinfo  {journal} {arXiv preprint arXiv:1802.06002}\ } (\bibinfo {year} {2018})}\BibitemShut {NoStop}%
\bibitem [{\citenamefont {Peruzzo}\ \emph {et~al.}(2014)\citenamefont {Peruzzo}, \citenamefont {McClean}, \citenamefont {Shadbolt}, \citenamefont {Yung}, \citenamefont {Zhou}, \citenamefont {Love}, \citenamefont {Aspuru-Guzik},\ and\ \citenamefont {O'brien}}]{peruzzo2014variational}%
  \BibitemOpen
  \bibfield  {author} {\bibinfo {author} {\bibfnamefont {A.}~\bibnamefont {Peruzzo}}, \bibinfo {author} {\bibfnamefont {J.}~\bibnamefont {McClean}}, \bibinfo {author} {\bibfnamefont {P.}~\bibnamefont {Shadbolt}}, \bibinfo {author} {\bibfnamefont {M.-H.}\ \bibnamefont {Yung}}, \bibinfo {author} {\bibfnamefont {X.-Q.}\ \bibnamefont {Zhou}}, \bibinfo {author} {\bibfnamefont {P.~J.}\ \bibnamefont {Love}}, \bibinfo {author} {\bibfnamefont {A.}~\bibnamefont {Aspuru-Guzik}},\ and\ \bibinfo {author} {\bibfnamefont {J.~L.}\ \bibnamefont {O'brien}},\ }\href@noop {} {\bibfield  {journal} {\bibinfo  {journal} {Nature communications}\ }\textbf {\bibinfo {volume} {5}},\ \bibinfo {pages} {4213} (\bibinfo {year} {2014})}\BibitemShut {NoStop}%
\bibitem [{\citenamefont {Gonthier}\ \emph {et~al.}(2022)\citenamefont {Gonthier}, \citenamefont {Radin}, \citenamefont {Buda}, \citenamefont {Doskocil}, \citenamefont {Abuan},\ and\ \citenamefont {Romero}}]{gonthier2022measurements}%
  \BibitemOpen
  \bibfield  {author} {\bibinfo {author} {\bibfnamefont {J.~F.}\ \bibnamefont {Gonthier}}, \bibinfo {author} {\bibfnamefont {M.~D.}\ \bibnamefont {Radin}}, \bibinfo {author} {\bibfnamefont {C.}~\bibnamefont {Buda}}, \bibinfo {author} {\bibfnamefont {E.~J.}\ \bibnamefont {Doskocil}}, \bibinfo {author} {\bibfnamefont {C.~M.}\ \bibnamefont {Abuan}},\ and\ \bibinfo {author} {\bibfnamefont {J.}~\bibnamefont {Romero}},\ }\href@noop {} {\bibfield  {journal} {\bibinfo  {journal} {Physical Review Research}\ }\textbf {\bibinfo {volume} {4}},\ \bibinfo {pages} {033154} (\bibinfo {year} {2022})}\BibitemShut {NoStop}%
\bibitem [{\citenamefont {McClean}\ \emph {et~al.}(2018)\citenamefont {McClean}, \citenamefont {Boixo}, \citenamefont {Smelyanskiy}, \citenamefont {Babbush},\ and\ \citenamefont {Neven}}]{mcclean2018barren}%
  \BibitemOpen
  \bibfield  {author} {\bibinfo {author} {\bibfnamefont {J.~R.}\ \bibnamefont {McClean}}, \bibinfo {author} {\bibfnamefont {S.}~\bibnamefont {Boixo}}, \bibinfo {author} {\bibfnamefont {V.~N.}\ \bibnamefont {Smelyanskiy}}, \bibinfo {author} {\bibfnamefont {R.}~\bibnamefont {Babbush}},\ and\ \bibinfo {author} {\bibfnamefont {H.}~\bibnamefont {Neven}},\ }\href@noop {} {\bibfield  {journal} {\bibinfo  {journal} {Nature communications}\ }\textbf {\bibinfo {volume} {9}},\ \bibinfo {pages} {4812} (\bibinfo {year} {2018})}\BibitemShut {NoStop}%
\bibitem [{\citenamefont {Hangleiter}\ and\ \citenamefont {Eisert}(2023)}]{hangleiter2023computational}%
  \BibitemOpen
  \bibfield  {author} {\bibinfo {author} {\bibfnamefont {D.}~\bibnamefont {Hangleiter}}\ and\ \bibinfo {author} {\bibfnamefont {J.}~\bibnamefont {Eisert}},\ }\href@noop {} {\bibfield  {journal} {\bibinfo  {journal} {Reviews of Modern Physics}\ }\textbf {\bibinfo {volume} {95}},\ \bibinfo {pages} {035001} (\bibinfo {year} {2023})}\BibitemShut {NoStop}%
\bibitem [{\citenamefont {Kanno}\ \emph {et~al.}(2023)\citenamefont {Kanno}, \citenamefont {Kohda}, \citenamefont {Imai}, \citenamefont {Koh}, \citenamefont {Mitarai}, \citenamefont {Mizukami},\ and\ \citenamefont {Nakagawa}}]{kanno2023quantum}%
  \BibitemOpen
  \bibfield  {author} {\bibinfo {author} {\bibfnamefont {K.}~\bibnamefont {Kanno}}, \bibinfo {author} {\bibfnamefont {M.}~\bibnamefont {Kohda}}, \bibinfo {author} {\bibfnamefont {R.}~\bibnamefont {Imai}}, \bibinfo {author} {\bibfnamefont {S.}~\bibnamefont {Koh}}, \bibinfo {author} {\bibfnamefont {K.}~\bibnamefont {Mitarai}}, \bibinfo {author} {\bibfnamefont {W.}~\bibnamefont {Mizukami}},\ and\ \bibinfo {author} {\bibfnamefont {Y.~O.}\ \bibnamefont {Nakagawa}},\ }\href@noop {} {\bibfield  {journal} {\bibinfo  {journal} {arXiv preprint arXiv:2302.11320}\ } (\bibinfo {year} {2023})}\BibitemShut {NoStop}%
\bibitem [{\citenamefont {Robledo-Moreno}\ \emph {et~al.}(2025)\citenamefont {Robledo-Moreno}, \citenamefont {Motta}, \citenamefont {Haas}, \citenamefont {Javadi-Abhari}, \citenamefont {Jurcevic}, \citenamefont {Kirby}, \citenamefont {Martiel}, \citenamefont {Sharma}, \citenamefont {Sharma}, \citenamefont {Shirakawa} \emph {et~al.}}]{robledo2025chemistry}%
  \BibitemOpen
  \bibfield  {author} {\bibinfo {author} {\bibfnamefont {J.}~\bibnamefont {Robledo-Moreno}}, \bibinfo {author} {\bibfnamefont {M.}~\bibnamefont {Motta}}, \bibinfo {author} {\bibfnamefont {H.}~\bibnamefont {Haas}}, \bibinfo {author} {\bibfnamefont {A.}~\bibnamefont {Javadi-Abhari}}, \bibinfo {author} {\bibfnamefont {P.}~\bibnamefont {Jurcevic}}, \bibinfo {author} {\bibfnamefont {W.}~\bibnamefont {Kirby}}, \bibinfo {author} {\bibfnamefont {S.}~\bibnamefont {Martiel}}, \bibinfo {author} {\bibfnamefont {K.}~\bibnamefont {Sharma}}, \bibinfo {author} {\bibfnamefont {S.}~\bibnamefont {Sharma}}, \bibinfo {author} {\bibfnamefont {T.}~\bibnamefont {Shirakawa}}, \emph {et~al.},\ }\href@noop {} {\bibfield  {journal} {\bibinfo  {journal} {Science Advances}\ }\textbf {\bibinfo {volume} {11}},\ \bibinfo {pages} {eadu9991} (\bibinfo {year} {2025})}\BibitemShut {NoStop}%
\bibitem [{\citenamefont {Layden}\ \emph {et~al.}(2023)\citenamefont {Layden}, \citenamefont {Mazzola}, \citenamefont {Mishmash}, \citenamefont {Motta}, \citenamefont {Wocjan}, \citenamefont {Kim},\ and\ \citenamefont {Sheldon}}]{layden2023quantum}%
  \BibitemOpen
  \bibfield  {author} {\bibinfo {author} {\bibfnamefont {D.}~\bibnamefont {Layden}}, \bibinfo {author} {\bibfnamefont {G.}~\bibnamefont {Mazzola}}, \bibinfo {author} {\bibfnamefont {R.~V.}\ \bibnamefont {Mishmash}}, \bibinfo {author} {\bibfnamefont {M.}~\bibnamefont {Motta}}, \bibinfo {author} {\bibfnamefont {P.}~\bibnamefont {Wocjan}}, \bibinfo {author} {\bibfnamefont {J.-S.}\ \bibnamefont {Kim}},\ and\ \bibinfo {author} {\bibfnamefont {S.}~\bibnamefont {Sheldon}},\ }\href@noop {} {\bibfield  {journal} {\bibinfo  {journal} {Nature}\ }\textbf {\bibinfo {volume} {619}},\ \bibinfo {pages} {282} (\bibinfo {year} {2023})}\BibitemShut {NoStop}%
\bibitem [{\citenamefont {Nakano}\ \emph {et~al.}(2026)\citenamefont {Nakano}, \citenamefont {Okada},\ and\ \citenamefont {Fujii}}]{nakano2025neural}%
  \BibitemOpen
  \bibfield  {author} {\bibinfo {author} {\bibfnamefont {Y.}~\bibnamefont {Nakano}}, \bibinfo {author} {\bibfnamefont {K.~N.}\ \bibnamefont {Okada}},\ and\ \bibinfo {author} {\bibfnamefont {K.}~\bibnamefont {Fujii}},\ }\href@noop {} {\bibfield  {journal} {\bibinfo  {journal} {PRX Quantum}\ }\textbf {\bibinfo {volume} {7}},\ \bibinfo {pages} {010338} (\bibinfo {year} {2026})}\BibitemShut {NoStop}%
\bibitem [{\citenamefont {Nakano}\ \emph {et~al.}(2024)\citenamefont {Nakano}, \citenamefont {Hakoshima}, \citenamefont {Mitarai},\ and\ \citenamefont {Fujii}}]{nakano2024markov}%
  \BibitemOpen
  \bibfield  {author} {\bibinfo {author} {\bibfnamefont {Y.}~\bibnamefont {Nakano}}, \bibinfo {author} {\bibfnamefont {H.}~\bibnamefont {Hakoshima}}, \bibinfo {author} {\bibfnamefont {K.}~\bibnamefont {Mitarai}},\ and\ \bibinfo {author} {\bibfnamefont {K.}~\bibnamefont {Fujii}},\ }\href@noop {} {\bibfield  {journal} {\bibinfo  {journal} {Physical Review Research}\ }\textbf {\bibinfo {volume} {6}},\ \bibinfo {pages} {033105} (\bibinfo {year} {2024})}\BibitemShut {NoStop}%
\bibitem [{\citenamefont {Lotshaw}\ \emph {et~al.}(2023)\citenamefont {Lotshaw}, \citenamefont {Siopsis}, \citenamefont {Ostrowski}, \citenamefont {Herrman}, \citenamefont {Alam}, \citenamefont {Powers},\ and\ \citenamefont {Humble}}]{lotshaw2023approximate}%
  \BibitemOpen
  \bibfield  {author} {\bibinfo {author} {\bibfnamefont {P.~C.}\ \bibnamefont {Lotshaw}}, \bibinfo {author} {\bibfnamefont {G.}~\bibnamefont {Siopsis}}, \bibinfo {author} {\bibfnamefont {J.}~\bibnamefont {Ostrowski}}, \bibinfo {author} {\bibfnamefont {R.}~\bibnamefont {Herrman}}, \bibinfo {author} {\bibfnamefont {R.}~\bibnamefont {Alam}}, \bibinfo {author} {\bibfnamefont {S.}~\bibnamefont {Powers}},\ and\ \bibinfo {author} {\bibfnamefont {T.~S.}\ \bibnamefont {Humble}},\ }\href@noop {} {\bibfield  {journal} {\bibinfo  {journal} {Physical Review A}\ }\textbf {\bibinfo {volume} {108}},\ \bibinfo {pages} {042411} (\bibinfo {year} {2023})}\BibitemShut {NoStop}%
\bibitem [{\citenamefont {D{\'\i}ez-Valle}\ \emph {et~al.}(2023)\citenamefont {D{\'\i}ez-Valle}, \citenamefont {Porras},\ and\ \citenamefont {Garc{\'\i}a-Ripoll}}]{diez2023quantum}%
  \BibitemOpen
  \bibfield  {author} {\bibinfo {author} {\bibfnamefont {P.}~\bibnamefont {D{\'\i}ez-Valle}}, \bibinfo {author} {\bibfnamefont {D.}~\bibnamefont {Porras}},\ and\ \bibinfo {author} {\bibfnamefont {J.~J.}\ \bibnamefont {Garc{\'\i}a-Ripoll}},\ }\href@noop {} {\bibfield  {journal} {\bibinfo  {journal} {Physical review letters}\ }\textbf {\bibinfo {volume} {130}},\ \bibinfo {pages} {050601} (\bibinfo {year} {2023})}\BibitemShut {NoStop}%
\bibitem [{\citenamefont {Germain}\ \emph {et~al.}(2015)\citenamefont {Germain}, \citenamefont {Gregor}, \citenamefont {Murray},\ and\ \citenamefont {Larochelle}}]{germain2015made}%
  \BibitemOpen
  \bibfield  {author} {\bibinfo {author} {\bibfnamefont {M.}~\bibnamefont {Germain}}, \bibinfo {author} {\bibfnamefont {K.}~\bibnamefont {Gregor}}, \bibinfo {author} {\bibfnamefont {I.}~\bibnamefont {Murray}},\ and\ \bibinfo {author} {\bibfnamefont {H.}~\bibnamefont {Larochelle}},\ }in\ \href@noop {} {\emph {\bibinfo {booktitle} {International conference on machine learning}}}\ (\bibinfo {organization} {PMLR},\ \bibinfo {year} {2015})\ pp.\ \bibinfo {pages} {881--889}\BibitemShut {NoStop}%
\bibitem [{\citenamefont {Uria}\ \emph {et~al.}(2016)\citenamefont {Uria}, \citenamefont {C{\^o}t{\'e}}, \citenamefont {Gregor}, \citenamefont {Murray},\ and\ \citenamefont {Larochelle}}]{uria2016neural}%
  \BibitemOpen
  \bibfield  {author} {\bibinfo {author} {\bibfnamefont {B.}~\bibnamefont {Uria}}, \bibinfo {author} {\bibfnamefont {M.-A.}\ \bibnamefont {C{\^o}t{\'e}}}, \bibinfo {author} {\bibfnamefont {K.}~\bibnamefont {Gregor}}, \bibinfo {author} {\bibfnamefont {I.}~\bibnamefont {Murray}},\ and\ \bibinfo {author} {\bibfnamefont {H.}~\bibnamefont {Larochelle}},\ }\href@noop {} {\bibfield  {journal} {\bibinfo  {journal} {Journal of Machine Learning Research}\ }\textbf {\bibinfo {volume} {17}},\ \bibinfo {pages} {1} (\bibinfo {year} {2016})}\BibitemShut {NoStop}%
\bibitem [{\citenamefont {Larochelle}\ and\ \citenamefont {Murray}(2011)}]{larochelle2011neural}%
  \BibitemOpen
  \bibfield  {author} {\bibinfo {author} {\bibfnamefont {H.}~\bibnamefont {Larochelle}}\ and\ \bibinfo {author} {\bibfnamefont {I.}~\bibnamefont {Murray}},\ }in\ \href@noop {} {\emph {\bibinfo {booktitle} {Proceedings of the fourteenth international conference on artificial intelligence and statistics}}}\ (\bibinfo {organization} {JMLR Workshop and Conference Proceedings},\ \bibinfo {year} {2011})\ pp.\ \bibinfo {pages} {29--37}\BibitemShut {NoStop}%
\bibitem [{\citenamefont {Neuhaus}\ and\ \citenamefont {Hager}(2003)}]{neuhaus20032d}%
  \BibitemOpen
  \bibfield  {author} {\bibinfo {author} {\bibfnamefont {T.}~\bibnamefont {Neuhaus}}\ and\ \bibinfo {author} {\bibfnamefont {J.~S.}\ \bibnamefont {Hager}},\ }\href@noop {} {\bibfield  {journal} {\bibinfo  {journal} {Journal of statistical physics}\ }\textbf {\bibinfo {volume} {113}},\ \bibinfo {pages} {47} (\bibinfo {year} {2003})}\BibitemShut {NoStop}%
\bibitem [{\citenamefont {Nu{\ss}baumer}\ \emph {et~al.}(2010)\citenamefont {Nu{\ss}baumer}, \citenamefont {Bittner},\ and\ \citenamefont {Janke}}]{nussbaumer2010free}%
  \BibitemOpen
  \bibfield  {author} {\bibinfo {author} {\bibfnamefont {A.}~\bibnamefont {Nu{\ss}baumer}}, \bibinfo {author} {\bibfnamefont {E.}~\bibnamefont {Bittner}},\ and\ \bibinfo {author} {\bibfnamefont {W.}~\bibnamefont {Janke}},\ }\href@noop {} {\bibfield  {journal} {\bibinfo  {journal} {Progress of Theoretical Physics Supplement}\ }\textbf {\bibinfo {volume} {184}},\ \bibinfo {pages} {400} (\bibinfo {year} {2010})}\BibitemShut {NoStop}%
\bibitem [{\citenamefont {Nau}\ \emph {et~al.}(2025)\citenamefont {Nau}, \citenamefont {Nutricati}, \citenamefont {Camino}, \citenamefont {Warburton},\ and\ \citenamefont {Maier}}]{nau2025quantum}%
  \BibitemOpen
  \bibfield  {author} {\bibinfo {author} {\bibfnamefont {M.~A.}\ \bibnamefont {Nau}}, \bibinfo {author} {\bibfnamefont {L.~A.}\ \bibnamefont {Nutricati}}, \bibinfo {author} {\bibfnamefont {B.}~\bibnamefont {Camino}}, \bibinfo {author} {\bibfnamefont {P.~A.}\ \bibnamefont {Warburton}},\ and\ \bibinfo {author} {\bibfnamefont {A.~K.}\ \bibnamefont {Maier}},\ }\href@noop {} {\bibfield  {journal} {\bibinfo  {journal} {Scientific Reports}\ }\textbf {\bibinfo {volume} {15}},\ \bibinfo {pages} {28937} (\bibinfo {year} {2025})}\BibitemShut {NoStop}%
\bibitem [{\citenamefont {M{\"u}cke}\ \emph {et~al.}(2023)\citenamefont {M{\"u}cke}, \citenamefont {Heese}, \citenamefont {M{\"u}ller}, \citenamefont {Wolter},\ and\ \citenamefont {Piatkowski}}]{mucke2023feature}%
  \BibitemOpen
  \bibfield  {author} {\bibinfo {author} {\bibfnamefont {S.}~\bibnamefont {M{\"u}cke}}, \bibinfo {author} {\bibfnamefont {R.}~\bibnamefont {Heese}}, \bibinfo {author} {\bibfnamefont {S.}~\bibnamefont {M{\"u}ller}}, \bibinfo {author} {\bibfnamefont {M.}~\bibnamefont {Wolter}},\ and\ \bibinfo {author} {\bibfnamefont {N.}~\bibnamefont {Piatkowski}},\ }\href@noop {} {\bibfield  {journal} {\bibinfo  {journal} {Quantum Machine Intelligence}\ }\textbf {\bibinfo {volume} {5}},\ \bibinfo {pages} {11} (\bibinfo {year} {2023})}\BibitemShut {NoStop}%
\bibitem [{\citenamefont {Hohenberg}\ and\ \citenamefont {Halperin}(1977)}]{hohenberg1977theory}%
  \BibitemOpen
  \bibfield  {author} {\bibinfo {author} {\bibfnamefont {P.~C.}\ \bibnamefont {Hohenberg}}\ and\ \bibinfo {author} {\bibfnamefont {B.~I.}\ \bibnamefont {Halperin}},\ }\href@noop {} {\bibfield  {journal} {\bibinfo  {journal} {Reviews of Modern Physics}\ }\textbf {\bibinfo {volume} {49}},\ \bibinfo {pages} {435} (\bibinfo {year} {1977})}\BibitemShut {NoStop}%
\bibitem [{\citenamefont {Kuchukova}\ \emph {et~al.}(2025)\citenamefont {Kuchukova}, \citenamefont {Pappik}, \citenamefont {Perkins},\ and\ \citenamefont {Yap}}]{kuchukova2025fast}%
  \BibitemOpen
  \bibfield  {author} {\bibinfo {author} {\bibfnamefont {A.}~\bibnamefont {Kuchukova}}, \bibinfo {author} {\bibfnamefont {M.}~\bibnamefont {Pappik}}, \bibinfo {author} {\bibfnamefont {W.}~\bibnamefont {Perkins}},\ and\ \bibinfo {author} {\bibfnamefont {C.}~\bibnamefont {Yap}},\ }\href@noop {} {\bibfield  {journal} {\bibinfo  {journal} {Random Structures \& Algorithms}\ }\textbf {\bibinfo {volume} {67}},\ \bibinfo {pages} {e70038} (\bibinfo {year} {2025})}\BibitemShut {NoStop}%
\bibitem [{\citenamefont {Hadfield}\ \emph {et~al.}(2019)\citenamefont {Hadfield}, \citenamefont {Wang}, \citenamefont {O'gorman}, \citenamefont {Rieffel}, \citenamefont {Venturelli},\ and\ \citenamefont {Biswas}}]{hadfield2019quantum}%
  \BibitemOpen
  \bibfield  {author} {\bibinfo {author} {\bibfnamefont {S.}~\bibnamefont {Hadfield}}, \bibinfo {author} {\bibfnamefont {Z.}~\bibnamefont {Wang}}, \bibinfo {author} {\bibfnamefont {B.}~\bibnamefont {O'gorman}}, \bibinfo {author} {\bibfnamefont {E.~G.}\ \bibnamefont {Rieffel}}, \bibinfo {author} {\bibfnamefont {D.}~\bibnamefont {Venturelli}},\ and\ \bibinfo {author} {\bibfnamefont {R.}~\bibnamefont {Biswas}},\ }\href@noop {} {\bibfield  {journal} {\bibinfo  {journal} {Algorithms}\ }\textbf {\bibinfo {volume} {12}},\ \bibinfo {pages} {34} (\bibinfo {year} {2019})}\BibitemShut {NoStop}%
\bibitem [{\citenamefont {Wang}\ \emph {et~al.}(2020)\citenamefont {Wang}, \citenamefont {Rubin}, \citenamefont {Dominy},\ and\ \citenamefont {Rieffel}}]{wang2020xy}%
  \BibitemOpen
  \bibfield  {author} {\bibinfo {author} {\bibfnamefont {Z.}~\bibnamefont {Wang}}, \bibinfo {author} {\bibfnamefont {N.~C.}\ \bibnamefont {Rubin}}, \bibinfo {author} {\bibfnamefont {J.~M.}\ \bibnamefont {Dominy}},\ and\ \bibinfo {author} {\bibfnamefont {E.~G.}\ \bibnamefont {Rieffel}},\ }\href@noop {} {\bibfield  {journal} {\bibinfo  {journal} {Physical Review A}\ }\textbf {\bibinfo {volume} {101}},\ \bibinfo {pages} {012320} (\bibinfo {year} {2020})}\BibitemShut {NoStop}%
\bibitem [{\citenamefont {Papamakarios}\ \emph {et~al.}(2017)\citenamefont {Papamakarios}, \citenamefont {Pavlakou},\ and\ \citenamefont {Murray}}]{papamakarios2017masked}%
  \BibitemOpen
  \bibfield  {author} {\bibinfo {author} {\bibfnamefont {G.}~\bibnamefont {Papamakarios}}, \bibinfo {author} {\bibfnamefont {T.}~\bibnamefont {Pavlakou}},\ and\ \bibinfo {author} {\bibfnamefont {I.}~\bibnamefont {Murray}},\ }\href@noop {} {\bibfield  {journal} {\bibinfo  {journal} {Advances in neural information processing systems}\ }\textbf {\bibinfo {volume} {30}} (\bibinfo {year} {2017})}\BibitemShut {NoStop}%
\bibitem [{\citenamefont {Metropolis}\ \emph {et~al.}(1953)\citenamefont {Metropolis}, \citenamefont {Rosenbluth}, \citenamefont {Rosenbluth}, \citenamefont {Teller},\ and\ \citenamefont {Teller}}]{metropolis1953equation}%
  \BibitemOpen
  \bibfield  {author} {\bibinfo {author} {\bibfnamefont {N.}~\bibnamefont {Metropolis}}, \bibinfo {author} {\bibfnamefont {A.~W.}\ \bibnamefont {Rosenbluth}}, \bibinfo {author} {\bibfnamefont {M.~N.}\ \bibnamefont {Rosenbluth}}, \bibinfo {author} {\bibfnamefont {A.~H.}\ \bibnamefont {Teller}},\ and\ \bibinfo {author} {\bibfnamefont {E.}~\bibnamefont {Teller}},\ }\href@noop {} {\bibfield  {journal} {\bibinfo  {journal} {The journal of chemical physics}\ }\textbf {\bibinfo {volume} {21}},\ \bibinfo {pages} {1087} (\bibinfo {year} {1953})}\BibitemShut {NoStop}%
\bibitem [{\citenamefont {Hastings}(1970)}]{hastings1970monte}%
  \BibitemOpen
  \bibfield  {author} {\bibinfo {author} {\bibfnamefont {W.~K.}\ \bibnamefont {Hastings}},\ }\href@noop {} {\bibfield  {journal} {\bibinfo  {journal} {Biometrika}\ }\textbf {\bibinfo {volume} {57}},\ \bibinfo {pages} {97} (\bibinfo {year} {1970})}\BibitemShut {NoStop}%
\bibitem [{\citenamefont {Kawasaki}(1966)}]{kawasaki1966diffusion}%
  \BibitemOpen
  \bibfield  {author} {\bibinfo {author} {\bibfnamefont {K.}~\bibnamefont {Kawasaki}},\ }\href@noop {} {\bibfield  {journal} {\bibinfo  {journal} {Physical Review}\ }\textbf {\bibinfo {volume} {145}},\ \bibinfo {pages} {224} (\bibinfo {year} {1966})}\BibitemShut {NoStop}%
\bibitem [{\citenamefont {Hukushima}\ and\ \citenamefont {Nemoto}(1996)}]{hukushima1996exchange}%
  \BibitemOpen
  \bibfield  {author} {\bibinfo {author} {\bibfnamefont {K.}~\bibnamefont {Hukushima}}\ and\ \bibinfo {author} {\bibfnamefont {K.}~\bibnamefont {Nemoto}},\ }\href@noop {} {\bibfield  {journal} {\bibinfo  {journal} {Journal of the Physical Society of Japan}\ }\textbf {\bibinfo {volume} {65}},\ \bibinfo {pages} {1604} (\bibinfo {year} {1996})}\BibitemShut {NoStop}%
\bibitem [{\citenamefont {Katzgraber}\ \emph {et~al.}(2006)\citenamefont {Katzgraber}, \citenamefont {K{\"o}rner},\ and\ \citenamefont {Young}}]{katzgraber2006universality}%
  \BibitemOpen
  \bibfield  {author} {\bibinfo {author} {\bibfnamefont {H.~G.}\ \bibnamefont {Katzgraber}}, \bibinfo {author} {\bibfnamefont {M.}~\bibnamefont {K{\"o}rner}},\ and\ \bibinfo {author} {\bibfnamefont {A.~P.}\ \bibnamefont {Young}},\ }\href@noop {} {\bibfield  {journal} {\bibinfo  {journal} {Physical Review B—Condensed Matter and Materials Physics}\ }\textbf {\bibinfo {volume} {73}},\ \bibinfo {pages} {224432} (\bibinfo {year} {2006})}\BibitemShut {NoStop}%
\bibitem [{\citenamefont {Peng}\ \emph {et~al.}(2005)\citenamefont {Peng}, \citenamefont {Long},\ and\ \citenamefont {Ding}}]{peng2005feature}%
  \BibitemOpen
  \bibfield  {author} {\bibinfo {author} {\bibfnamefont {H.}~\bibnamefont {Peng}}, \bibinfo {author} {\bibfnamefont {F.}~\bibnamefont {Long}},\ and\ \bibinfo {author} {\bibfnamefont {C.}~\bibnamefont {Ding}},\ }\href@noop {} {\bibfield  {journal} {\bibinfo  {journal} {IEEE Transactions on pattern analysis and machine intelligence}\ }\textbf {\bibinfo {volume} {27}},\ \bibinfo {pages} {1226} (\bibinfo {year} {2005})}\BibitemShut {NoStop}%
\end{thebibliography}
%

\end{document}